\documentclass[11pt,a4paper]{article}

\usepackage[T1]{fontenc}
\usepackage{lmodern}
\usepackage{microtype}

\usepackage{amsmath,amsthm,amssymb,mathtools}
\usepackage{bbm}        
\usepackage{accents}    
\usepackage{tensor}
\usepackage{stmaryrd}

\usepackage{geometry}
\usepackage{parskip}
\usepackage{needspace}
\usepackage{calc}
\usepackage{setspace}

\usepackage{titlesec}
\titlespacing*{\section}   {0pt}{1.5em plus 1ex minus .2ex}{0.3em plus .2ex}
\titlespacing*{\subsection}{0pt}{1.2em plus 1ex minus .2ex}{0.2em plus .2ex}
\titlespacing*{\subsubsection}{0pt}{1.2em plus 1ex minus .2ex}{0.2em plus .2ex}

\usepackage{titling}
\posttitle{\par\end{center}\vspace{.5em}}

\usepackage{xcolor}
\definecolor{mmablue}  {RGB}{  0, 70, 160}
\definecolor{mmaorange}{RGB}{180, 90,   0}
\definecolor{mmagray}  {RGB}{110,110, 110}
\definecolor{lightgray}{gray}{0.96}

\usepackage{hyperref}
\hypersetup{
    hidelinks,
    breaklinks = true,
    colorlinks = true,
    linkcolor  = blue,
    urlcolor   = black!30!red,
    citecolor  = black!30!blue,
}
\usepackage[noabbrev]{cleveref}
\crefname{subsection}   {subsection}   {subsections}
\crefname{subsubsection}{subsubsection}{subsubsections}

\usepackage{enumitem}

\usepackage[sort&compress]{natbib}

\usepackage{tcolorbox}
\tcbuselibrary{skins, breakable, theorems}

\newtcbtheorem{example}{Example}{
    blanker,
    breakable,
    before={\needspace{6\baselineskip}}, 
    left=15pt, top=5pt, bottom=5pt,
    left skip=10pt,
    borderline west={3pt}{0pt}{gray!50},
    coltitle=black,
    fonttitle=\bfseries,
    terminator sign={:},
    attach title to upper={\par\vspace{-3ex}\setlength{\parskip}{\baselineskip}},
}{example}

\makeatletter
\crefname{tcb@cnt@example}{example}{examples}
\crefname{tcb@cnt@example}{Example}{Examples}
\makeatother

\usepackage{xstring}

\newif\ifmmaInList
\newcommand{\mmacont}{\textcolor{mmagray}{\ensuremath{\hookrightarrow}}\hspace{4pt}}

\newcounter{mmaCounter}
\newcounter{mmaFirstIn}
\newcounter{mmaOutLine}
\newcommand{\mmaSuppressedList}{,}
\newcommand{\mmaCurrentInputLine}{1}

\newcommand{\mmaSup}{%
  \xdef\mmaSuppressedList{\mmaSuppressedList\mmaCurrentInputLine,}%
}

\newcommand{\mmaCheckSuppressed}{%
  \edef\mmaCheckNeedle{,\arabic{mmaOutLine},}%
  \expandafter\IfSubStr\expandafter{\mmaSuppressedList}{\mmaCheckNeedle}%
    {\mmaInListtrue}%
    {\mmaInListfalse}%
}

\newcommand{\mmaSkipSuppressed}{%
  \mmaCheckSuppressed%
  \ifmmaInList%
    \stepcounter{mmaCounter}%
    \stepcounter{mmaOutLine}%
    \mmaSkipSuppressed%
  \fi%
}

\newtcolorbox{mmaIn}{
  enhanced,
  before skip=0.4em,   
  boxrule=0.4pt,
  colback=mmablue!4!white,
  colframe=mmablue!80!black,
  sharp corners,
  fontupper=\ttfamily\small\upshape,
  left=41pt,
  top=1pt, bottom=2pt,
  before upper={%
    \parindent=0pt%
    \setcounter{mmaFirstIn}{\value{mmaCounter}}%
    \gdef\mmaSuppressedList{,}%
    \gdef\mmaCurrentInputLine{1}%
    \setlength{\baselineskip}{1.1\baselineskip}%
    \noindent\llap{\scriptsize\ttfamily\bfseries\color{mmablue}In[\arabic{mmaCounter}]:=\hspace{5pt}}%
    \def\\{\par\noindent\llap{\mmacont}}%
    \renewcommand{\newline}{%
      \stepcounter{mmaCounter}%
      \xdef\mmaCurrentInputLine{\the\numexpr\mmaCurrentInputLine+1\relax}%
      \par\noindent}%
  },
  after=\stepcounter{mmaCounter}
}

\newcounter{mmaSavedCounter}

\newenvironment{mmaOut}{%
  \setcounter{mmaSavedCounter}{\value{mmaCounter}}%
  \begin{mmaOutBase}%
  \parindent=0pt%
  \setcounter{mmaCounter}{\value{mmaFirstIn}}%
  \setcounter{mmaOutLine}{1}%
  \setlength{\baselineskip}{1.1\baselineskip}%
  \upshape%
  \mmaSkipSuppressed%
  \noindent
  \llap{\scriptsize\ttfamily\bfseries\color{mmaorange}Out[\arabic{mmaCounter}]=\hspace{-1pt}}%
  \expandafter\def\csname newline \endcsname{\par\noindent}%
  \def\\{%
    \par\vspace{4pt}%
    \stepcounter{mmaOutLine}%
    \stepcounter{mmaCounter}%
    \mmaSkipSuppressed%
    \noindent\llap{\scriptsize\ttfamily\bfseries\color{mmaorange}%
      Out[\arabic{mmaCounter}]=\hspace{-1pt}}}%
}{%
  \xdef\mmaAfterOut{\arabic{mmaCounter}}%
  \end{mmaOutBase}%
  \setcounter{mmaCounter}{\mmaAfterOut}%
  \stepcounter{mmaCounter}%
}

\newtcolorbox{mmaOutBase}{
  enhanced,
  before skip=0.4em,   
  boxrule=0.4pt,
  colback=mmaorange!4!white,
  colframe=mmaorange!80!black,
  sharp corners,
  fontupper=\ttfamily\small,
  left=41pt, % Changed from 35pt to match mmaIn
  top=1pt, bottom=2pt,
  before upper={\parindent=0pt},
}

\newif\ifmmaFirstLabel

\newcommand{\mmaErr}[2]{%
  \par\vspace{2pt}%
  \noindent\hspace*{-22pt}% Pulls the text into the "Out" label zone
  \begin{minipage}{\linewidth+32pt}\setstretch{0.75}% Expands width to prevent premature wrapping
    \textcolor{red!81!black}{\ttfamily\scriptsize\bfseries #1::}%
    \textcolor{red!11!black}{\ttfamily\scriptsize\ #2}%
  \end{minipage}%
  \par\vspace{2pt}\noindent\ignorespaces%
}

\renewenvironment{mmaOut}[1][standard]{%
  \setcounter{mmaSavedCounter}{\value{mmaCounter}}%
  \begin{mmaOutBase}%
  \setcounter{mmaCounter}{\value{mmaFirstIn}}%
  \setcounter{mmaOutLine}{1}%
  \setlength{\baselineskip}{1.0\baselineskip}% 
  \upshape%
  \mmaSkipSuppressed%
  \def\temp{#1}\def\optNoLabel{empty}%
  \ifx\temp\optNoLabel%
    \mmaFirstLabeltrue%
    \noindent\ignorespaces%
  \else%
    \mmaFirstLabelfalse%
    \leavevmode\llap{\scriptsize\ttfamily\bfseries\color{mmaorange}Out[\arabic{mmaCounter}]=\hspace{5pt}}\ignorespaces%
  \fi%
  \def\\{%
    \par\vspace{2pt}% Reduced from 4pt to keep math lines closer
    \ifmmaFirstLabel%
      \mmaFirstLabelfalse%
    \else%
      \stepcounter{mmaOutLine}%
      \stepcounter{mmaCounter}%
      \mmaSkipSuppressed%
    \fi%
    \noindent\llap{\scriptsize\ttfamily\bfseries\color{mmaorange}Out[\arabic{mmaCounter}]=\hspace{5pt}}\ignorespaces%
  }%
}{%
  \xdef\mmaAfterOut{\arabic{mmaCounter}}%
  \end{mmaOutBase}%
  \setcounter{mmaCounter}{\mmaAfterOut}%
  \stepcounter{mmaCounter}%
}

\newcommand{\mmaGraphicOut}[2][]{%
  \leavevmode\raisebox{\dimexpr-\height+\baselineskip}{\includegraphics[#1]{#2}}%
}

\newcommand{\mmaStep}{%
  \stepcounter{mmaCounter}%
  \xdef\mmaLastIn{\arabic{mmaCounter}}%
}

\theoremstyle{plain}

\theoremstyle{plain}
\newtheorem{definition}{Definition}[section] 
\newtheorem{remark}[definition]{Remark}      

\newenvironment{remarks}{%
    \par\medskip
    \noindent\textbf{Remarks.}% 
    \begin{enumerate}[label=\textbf{\upshape\thesection.\arabic*.}]
    \itshape
    \setcounter{enumi}{\value{definition}} 
}{%
    \setcounter{definition}{\value{enumi}} 
    \end{enumerate}
    \medskip
}

\newcommand{\dd}  {\mathbbm{d}}
\newcommand{\ii}  {\mathbbm{i}}
\newcommand{\jj}  {\mathbbm{j}}
\newcommand{\LL}  {\mathbbm{L}}
\newcommand{\QQ}  {\mathbbm{Q}}
\newcommand{\XX}  {\mathbb{X}}

\newcommand{\HH}  {\mathbb{H}}
\newcommand{\R}   {\mathbb{R}}
\renewcommand{\SS}{\mathbb{S}}
\renewcommand{\d} {\mathrm{d}}
\renewcommand{\L} {\mathcal{L}}
\newcommand{\F}   {\mathcal{F}}
\newcommand{\M}   {\mathcal{M}}
\newcommand{\vol} {\mathrm{vol}}
\newcommand{\Sol} {\mathrm{Sol}}
\newcommand{\dbtilde}[1]{\accentset{\approx}{#1}}

\def\wwedgee{{\setbox0\hbox{\ensuremath{\mathrel{\wedge}}}%
    \rlap{\hbox to \wd0{\hss\,\ensuremath\wedge\hss}}\box0}}
\newcommand{\wwedge}{\mathrel{\wwedgee}}

\newcommand{\xAct}      {\texttt{\textup{xAct}}\xspace}
\newcommand{\xCore}     {\texttt{\textup{xCore}}\xspace}
\newcommand{\xCoba}     {\texttt{\textup{xCoba}}\xspace}
\newcommand{\xPerm}     {\texttt{\textup{xPerm}}\xspace}
\newcommand{\xPert}     {\texttt{\textup{xPert}}\xspace}
\newcommand{\xTensor}   {\texttt{\textup{xTensor}}\xspace}
\newcommand{\xTerior}   {\texttt{\textup{xTerior}}\xspace}
\newcommand{\xCPS}      {\texttt{\textup{xCPS}}\xspace}
\newcommand{\TInvar}      {\texttt{\textup{TInvar}}\xspace}
\newcommand{\TexAct}    {\texttt{\textup{TexAct}}\xspace}
\newcommand{\mathematica}{\textsc{\textup{Mathematica}}\xspace}
\newcommand{\x}[1]      {\texttt{\textup{#1}}}

\usepackage{todonotes}
\usepackage{xspace}

\title{%
    \LARGE\bfseries
    \xCPS: an \xAct package for covariant phase space, Noether charges, and entropy
}
\author{%
    Juan Margalef--Bentabol\\[4pt]
    {\small\itshape Département de mathématiques et de statistique, Université de Montréal, Canada}
}
\date{}

\begin{document}
\maketitle
\vspace{-1.5em}   

%/=====================\%
% Keywords
%\=====================/%
%\noindent\textbf{Keywords:} Tensor computer algebra, \mathematica, Covariant Phase Space, Variational bicomplex, Noether symmetries, Noether charges, \xAct
%/=====================\%
% Abstract
%\=====================/%
\begin{abstract}
\vspace{-0.9em} 

\noindent
\xCPS is an \xAct tensor algebra package for symbolic computations within the covariant phase space formalism of field theories. From a generic Lagrangian, \xCPS automates the derivation of equations of motion and symplectic currents. It systematically determines whether an infinitesimal transformation in the space of fields is a Noether symmetry and computes the associated Noether charge. Additionally, \xCPS can in many cases determine whether a tensorial expression is a divergence and, if so, find its divergence potential. By implementing vertical exterior calculus through a graded, supercommutative wedge product and vertical operators, the package enables efficient computations in gauge theories and higher-derivative models of gravity, including the derivation of thermodynamic quantities like Wald's entropy. \xCPS is open-source under the GPL license and available at \url{https://github.com/juanmargalef}.
\end{abstract}

%%%%%%%%%%%%%%%%%%%%%%%%%%%%%%%%%%%%%%%%%%%%%%%%%%%%%%%%%%%%%%%%%%%%%%%%%%%%%%%%%%%%%%%%%%
%%%%%%%%%%%%%%%%%%%%%%%%%%%%%%%%%%%%%%%%%%%%%%%%%%%%%%%%%%%%%%%%%%%%%%%%%%%%%%%%%%%%%%%%%%
%%%%%%%%%%%%%%%%%%%%%%%%%%%%%%%%%%%%%%%%%%%%%%%%%%%%%%%%%%%%%%%%%%%%%%%%%%%%%%%%%%%%%%%%%%

%/=====================\%
\section{Introduction}
%\=====================/%
To extract physical information from a local field theory, one can adopt either the canonical or the covariant framework. Although closely related, they are conceptually different. The former splits geometric objects into tangent (spatial) and normal (temporal) parts, specifies an initial state, and evolves it forward in time \cite{margalef2018towards}. Its main advantages are the dynamical perspective and the existence of universal structures like the symplectic form of a cotangent bundle, which are a natural starting point for numerical methods and quantization. However, this approach breaks certain symmetries like covariance, and complicates the study of asymptotic regions.

The covariant phase space (CPS) approach, by contrast, treats fields over the entire spacetime and, under mild conditions, is fully equivalent to the canonical one \cite{margalef2022proof}. Symmetries remain manifest, null infinity is accessible, conserved Noether quantities can be computed systematically, and higher-derivative theories are on equal footing with first-order ones. The main difficulty is the absence of canonical structures in the spaces involved. To obtain a symplectic structure (essential, among other things, for black hole thermodynamics and quantization), one must fix an action and rely on the CPS formalism \cite{Crnkovic1986,zuckerman1987action}, which has been extended to include boundaries \cite{margalef2021geometric} and successfully applied to several theories of gravity \cite{barbero2021covariant,barbero2021palatini,barbero2022shell}.

The CPS formalism has proven especially powerful for black hole thermodynamics \cite{iyer1994some,lee1990local,wald1993black,wald2000general} and the study of asymptotic symmetries \cite{bondi1962gravitational,sachs1962asymptotic}. Interest has been renewed recently by gravitational wave astronomy and the discovery of infinitely many conserved charges associated with asymptotic symmetry groups \cite{cachazo2014evidence,campiglia2014asymptotic,barnich2011bms,freidel2021weyl}. Despite its conceptual elegance, manual calculations quickly become intractable for theories involving higher-derivative terms or complex gauge structures. However, no general computational implementation of the CPS formalism exists.

To fill this gap, we introduce \xCPS, a \mathematica package designed to automate these calculations efficiently while preserving the geometric nature of the formalism. Built on the \xAct tensor algebra bundle \cite{xPerm}, \xCPS provides automated tools for variational calculus (equations of motion and symplectic currents), Noether symmetry testing, and the computation of Noether charges. It also includes algorithms for divergence detection and heuristic potential-finding. The package allows users to work with concrete Lagrangians (e.g., gravity or electromagnetism) or with \textbf{unspecified} Lagrangians depending on tensorial fields $\{\Phi^i\}_{i}$ and finitely many of their derivatives. By managing the intricate index manipulations and super-commutativity inherent in the CPS formalism, \xCPS allows researchers to focus on physics rather than algebraic bookkeeping. The following representative outputs illustrate the package's core capabilities (see the companion executable \cite{xCPS}):
\begin{enumerate}
    \item  The first variation of a generic Lagrangian depending on a scalar field and its first derivative:
\end{enumerate}
\begin{mmaIn}
    Lscalar = $\sqrt{-\tilde{\tilde{g}}}$ L[Phi,NablaPhi];\mmaSup\newline
    FirstVariation[Phi][Lscalar] // TotalDerivativeToCovD // Simplification
\end{mmaIn}
\begin{mmaOut}
    $  \sqrt{-\tilde{\tilde{g}}} \bigl( (\frac{\partial L}{\partial \Phi}) - \nabla_{a}(\frac{\partial L}{\partial \nabla \Phi})^{a}\bigr)(\dd \Phi)+\nabla_a\bigl[\sqrt{-\tilde{\tilde{g}}}(\dd \Phi) (\frac{\partial L}{\partial \nabla \Phi})^{a}\bigr] $
\end{mmaOut}
\begin{enumerate}[resume]
    \item The symplectic current of general relativity (GR) and its Wald entropy:
\end{enumerate} 
\begin{mmaIn}
    LGR = $\sqrt{-\tilde{\tilde{g}}}$ RicciScalarNabla[];\mmaSup\newline
    SymplecticCurrent[][LGR] // ContractMetric // Simplification\newline
EOM[RiemannNabla][LGR] // Simplification
\end{mmaIn} 
\begin{mmaOut}
    $\tfrac{1}{2}\sqrt{-\tilde{\tilde{g}}}(n_{{}_\nabla})^{a} \bigl\{(\dd g)_{a}{}^{b} \wwedge \nabla_{b}(\dd g)^{c}{}_{c} -  (\dd g)^{b}{}_{b} \wwedge \nabla_{a}(\dd g)^{c}{}_{c} + (\dd g)^{b}{}_{b} \wwedge \nabla_{c}(\dd g)_{a}{}^{c}$\newline$\mbox{}\hspace{14ex}+ (\dd g)^{bc} \wwedge \nabla_{a}(\dd g)_{bc} - 2 (\dd g)^{bc} \wwedge \nabla_{c}(\dd g)_{ab}\bigr\}$\\
    $\tfrac{1}{2}\sqrt{-\tilde{\tilde{g}}} (- g^{ad} g^{bc} + g^{ac} g^{bd})$
\end{mmaOut}

\begin{enumerate}[resume]
    \item The equation of motion of a generic $F(\mathrm{Riemann})$ theory:
\end{enumerate} 
\begin{mmaIn}
    LfRiem = $\sqrt{-\tilde{\tilde{g}}}$ F[RiemannNabla];\mmaSup\newline
    EOM[g][LfRiem] // ContractMetric // Simplification
\end{mmaIn}
\begin{mmaOut}
    $\tfrac{1}{2} \sqrt{-\tilde{\tilde{g}}}\bigl\{F\bigl[Riem\bigr] g^{ab}+(\frac{\partial \x{F}}{\partial Riem})^{bcde} Riem^{a}{}_{cde}+(\frac{\partial \x{F}}{\partial Riem})^{acde} Riem^{b}{}_{cde}$\newline$\mbox{}\hspace{10ex}-2\nabla_{c}\nabla_{d}(\frac{\partial \x{F}}{\partial Riem})^{acbd}-2\nabla_{d}\nabla_{c}(\frac{\partial \x{F}}{\partial Riem})^{acbd}\bigr\}$
\end{mmaOut}
\begin{enumerate}[resume]
    \item Wald's entropy of a generic $F(\mathrm{Riemann},\nabla\mathrm{Riemann})$ theory:
\end{enumerate} 
\begin{mmaIn}
    LfRiem2 = Sqrt[-Detg[]] F[RiemannNabla, NablaRiemannNabla];\mmaSup\newline
    EOM[RiemannNabla][LfRiem2] // ContractMetric // Simplification
\end{mmaIn}
\begin{mmaOut}
    $\sqrt{-\tilde{\tilde{g}}} ((\frac{\partial F}{\partial Riem})^{abcd} -  \nabla_{e}(\frac{\partial F}{\partial \nabla Riem})^{abcde})$
\end{mmaOut}

\begin{enumerate}[resume]
    \item Check the diff-invariance of GR and compute its conserved Noether charge:
\end{enumerate} 
\begin{mmaIn}
    DiffAction = VVFFromLieD[xi][g]\newline
    NoetherSymmetryQ[DiffAction][g][LGR] // SortCovDs // $\cdots{}$ // Simplification\newline
    NoetherCurrent[DiffAction][g][LGR,···] (*\,We will explain the dots later\,*)
\end{mmaIn}
\begin{mmaOut}
    $(\frac{\delta}{\delta g})^{ab} \mathcal{L}_\xi g_{ab}$\\
    True\\
    $\sqrt{-\tilde{\tilde{g}}}(n_{{}_\nabla})^{a} (R \,\xi_{a} - 2 Ric_{ab}\, \xi^{b})$ 
\end{mmaOut}
\begin{enumerate}[resume]
    \item Find the gradient potential of an expression:
\end{enumerate} 
\begin{mmaIn}
   expr = Phi[]Nabla[-a]@Nabla[-c]@xi[b] + Nabla[-a]@Phi[]Nabla[-c]@xi[b]\newline
FindPotentialGradient[Nabla][expr]
\end{mmaIn}
\begin{mmaOut}
    $\Phi \nabla_{a}\nabla_{c}\xi^{b} + \nabla_{a}\Phi \nabla_{c}\xi^{b}$\\
    $(n_{{}_\nabla})_{a} \Phi \nabla_{c}\xi^{b}$
\end{mmaOut}

This paper is organized as follows. \Cref{sec:theory} provides a brief overview of the theory underlying \xCPS while \cref{sec: xact} briefly introduces \xAct. \Cref{sec: functionalities} is devoted to explaining the core ideas of the package that are later applied in \cref{sec: variational calculus}, where we introduce the variational calculus implementation. In \cref{sec: final examples} some detailed examples are provided to showcase the full power of the package. Finally, \cref{sec: conclusions} presents our conclusions. Readers interested in applications may go directly to \cref{sec: variational calculus}, while those interested in the field theory examples may go straight to \cref{sec: final examples}.

%%%%%%%%%%%%%%%%%%%%%%%%%%%%%%%%%%%%%%%%%%%%%%%%%%%%%%%%%%%%%%%%%%%%%%%%%%%%%%%%%%%%%%%%%%
%%%%%%%%%%%%%%%%%%%%%%%%%%%%%%%%%%%%%%%%%%%%%%%%%%%%%%%%%%%%%%%%%%%%%%%%%%%%%%%%%%%%%%%%%%
%%%%%%%%%%%%%%%%%%%%%%%%%%%%%%%%%%%%%%%%%%%%%%%%%%%%%%%%%%%%%%%%%%%%%%%%%%%%%%%%%%%%%%%%%%

%/=====================\%
\section{Mathematical background}\label{sec:theory}
%\=====================/%
This section summarizes the mathematical foundation underlying \xCPS. We consider an $n$-manifold without boundary $\M$ and a space of tensorial fields $\F$ over $\M$. Both spaces are naturally endowed with an exterior derivative: $\d$ for $\M$ and $\dd$ for $\F$, (horizontal and vertical directions of the variational bicomplex \cite{anderson1989variational}). Following the standard physics convention, we assume they commute.

%/=-=-=-=-=-=-=\%
\subsection{First variation}
%\=-=-=-=-=-=-=/%
Consider a local Lagrangian $L:\F\to \Omega^n(\M)$, i.e., a top-form depending on the fields of $\F$ and finitely many of their derivatives. Equivalently, $L\in\Omega^{(n,0)}(\M\times\F)$ is a bi-graded $(n,0)$-form over $\M\times\F$ (see \cite{anderson1989variational,margalef2021geometric}). The fundamental CPS identity is the canonical decomposition of the first variation:
\begin{equation}\label{eq: first variation}
\dd L = \mathrm{EOM}_i \dd \Phi^i + \d\Theta
\end{equation}
where $\mathrm{EOM}_i\in\Omega^{(n,0)}(\M\times\F)$ is the Euler-Lagrange (EL) form whose kernel provides the solution to the equations of motion\footnote{To avoid cluttering the notation, we assume that $\Phi^i$ has horizontal degree $k=0$, though it can have any index structure. If $\Phi^i$ is a $k$-form with $k\neq0$, we have $\mathrm{EOM}_i\in\Omega^{(n-k,0)}(\M\times\F)$ and $\dd L = \mathrm{EOM}_i \wedge \dd \Phi^i + \d\Theta$, with the remaining  arguments unchanged.} $\mathrm{EOM}_i=0$ associated with the dynamical field $\Phi^i\in\F$; $\Theta\in\Omega^{(n-1,1)}(\M\times\F)$ is a symplectic potential; and $\dd \Phi^i$ acts as an infinitesimal variation of the field $\Phi^i$, analogous to $\d x, \d y, \d z$ in $\R^3$. The global decomposition \eqref{eq: first variation} holds under mild regularity conditions \cite{margalef2021geometric,anderson1989variational} and is achieved in practice by repeatedly applying the Leibniz rule to remove derivatives from $\dd \Phi^i$.
\begin{remark}
 In the physics literature, the equations of motion are often denoted as $\mathrm{EOM}_i=:\frac{\delta L}{\delta \Phi^i}$.
\end{remark}

\begin{example}{}{ first variation scalar}
Given a fixed metric $g$, consider the Lagrangian of a scalar field theory given by 
\begin{equation}\label{eq: Lagrangian scalar field}
    L_{\mathrm{scl}}(\phi)=\frac{1}{2}\nabla_a\phi\nabla^a\phi\,\vol_g
\end{equation}
where $\vol_g$ is the metric volume form. Its vertical exterior derivative is
\begin{equation}\label{eq: first variation scalar field 1}
\dd L_{\mathrm{scl}}=\frac{1}{2}\nabla_a\dd\phi\nabla^a\phi\,\vol_g+\frac{1}{2}\nabla_a\phi\nabla^a\dd\phi\,\vol_g=\nabla_a\dd\phi\nabla^a\phi\,\vol_g
\end{equation}
where we have used that $\dd g_{ab}=0$. Now, using $\nabla g_{ab}=0$ and the Leibniz rule leads to:
\begin{equation}\label{eq: first variation scalar field 2}
\dd L_{\mathrm{scl}}=\nabla_a(\dd\phi\nabla^a\phi)\vol_g-\dd\phi\nabla_a\nabla^a\phi\,\vol_g=\mathrm{EOM}_{\mathrm{scl}}\,\dd\phi+\mathrm{div}(\vec{U}_{\mathrm{scl}})\vol_g
\end{equation}
where $U_{\mathrm{scl}}^a=\dd\phi\nabla^a\phi$. The last term of \eqref{eq: first variation scalar field 2} can be rewritten using $\L_{\vec{U}}\vol_g=\mathrm{div}(\vec{U})\vol_g$, $\d\vol_g=0$, and  $\L_{\vec{U}}=\mathcal{\iota}_{\vec{U}}\d+\d\mathcal{\iota}_{\vec{U}}$ ($\mathcal{\iota}:\Omega^{k}(M)\to\Omega^{k-1}(M)$ is the interior product). Thus, the first variation takes the form of \cref{eq: first variation} with $\mathrm{EOM}_{\mathrm{scl}}=-\nabla_a\nabla^a\phi\vol_g \in\Omega^{(n,0)}(\M\times\F)$ the EL form of $\phi$ (whose kernel is the solutions to the wave equation $\nabla_a\nabla^a \phi=0$) and $\Theta_{\mathrm{scl}}=\iota_{\vec{U}_{\mathrm{scl}}}\vol_g\in\Omega^{(n-1,1)}(\M\times\F)$ a symplectic potential.
\end{example}

%/=-=-=-=-=-=-=\%
\subsection{Symplectic form}
%\=-=-=-=-=-=-=/%
For a given symplectic potential $\Theta$, we define its symplectic current as
\begin{equation}\label{eq: omega=dlTheta}
\omega=\dd\Theta\in\Omega^{(n-1,2)}(\M\times\F)
\end{equation}
Integrating the horizontal part of $\omega$ over an $(n-1)$-dimensional submanifold $\Sigma\subset \M$ yields a $2$-form on $\F$. The natural choice is to take $\Sigma$ a Cauchy surface. We then define the (pre)symplectic form

\begin{equation}\label{eq: Omega=int omega}
\Omega_\Sigma=\int_\Sigma \imath^*\omega\in\Omega^{2}(\F)
\end{equation}
where $\imath:\Sigma\hookrightarrow \M$ is the inclusion (though it is customary to omit this pullback, we retain it to avoid common misunderstandings). Clearly $\Omega_\Sigma$ is a closed $2$-form ($\dd^2=0$), though we cannot generally guarantee its non-degeneracy.

\begin{example}{}{}
From the Lagrangian \eqref{eq: Lagrangian scalar field}, we obtained $\Theta_{\mathrm{scl}}=\iota_{\vec{U}_{\mathrm{scl}}}\vol_g$ with $U_{\mathrm{scl}}^a=\dd\phi\nabla^a\phi$. This leads to 
\[\omega_{\mathrm{scl}}=\dd\Theta_{\mathrm{scl}}=\iota_{\dd\vec{U}_{\mathrm{scl}}}\vol_g\]
where $\dd U_{\mathrm{scl}}^a=\dd^2\phi\nabla^a\phi-\dd\phi\wwedge\nabla^a\dd\phi=-\dd\phi\wwedge\nabla^a\dd\phi$. Here we use the natural wedge product $\wwedge$ in the space of forms over $\F$, $\dd^2=0$, and $\dd g_{ab}=0$ since $g$ is fixed.

Integrating this current to obtain the symplectic form requires the future-directed, unit normal vector field $\vec{n}$ to the chosen Cauchy surface $\Sigma$ (in abstract-index notation: $n^a$). Over $\Sigma$, and with a slight abuse of notation, we have $\vol_g=-n\wedge\vol_\gamma$, where $n$ is the $1$-form metrically equivalent to $\vec{n}$ (with indices: $n_a$) and $\vol_\gamma$ is the volume metric form of the induced pullback metric $\gamma=\imath^*g$. This leads to the symplectic form:
\[(\Omega_{\mathrm{scl}})_\Sigma=\int_\Sigma\imath^*\iota_{\dd\vec{U}_{\mathrm{scl}}}\vol_g=\int_\Sigma(- n_a\dd\vec{U}_{\mathrm{scl}}^a)\vol_\gamma=\int_\Sigma \dd(\imath^*\phi)\wwedge\dd\imath^*(\L_{\vec{n}}\phi)\vol_\gamma\]
Roughly speaking, $\imath^*\phi$ and $\imath^*(\L_{\vec{n}}\phi)$ correspond to the position and momenta in the canonical formalism \cite{margalef2022proof}. Thus, as expected, $(\Omega_{\mathrm{scl}})_\Sigma$ corresponds to the canonical symplectic form.
%\[\Omega_\Sigma=\int_\Sigma \dd q\wwedge\dd p\,\vol_\gamma\]
\end{example}

%/=-=-=-=-=-=-=\%
\subsection{Ambiguities}\label{subsection: Ambiguities}
%\=-=-=-=-=-=-=/%
The preceding constructions involve several choices. We now verify that the physically relevant objects are independent of them. First, ``the'' potential $\Theta$ is not well-defined: any potential $\Theta'=\Theta+\alpha$ (for a horizontally-closed $\alpha\in\Omega^{(n-1,1)}(\M\times\F)$) is equally valid. Because the vertical degree is non-zero, the generalized Poincaré lemma implies $\alpha$ is exact ($\alpha=\d Z$) \cite{wald1990identically}. Thus, while $\Theta$ is ambiguous, its cohomology class $[\Theta]$ is not.% This is a common occurrence: the natural playground is the horizontal cohomology. %For precaution, it is customary to consider the symplectic potential $\Theta'=\Theta+\d Z$ since this cover all possible choices of potential.

A second source of ambiguity arises from the Lagrangian. Since two Lagrangians equal up to a total derivative define the same action, $\SS_L=\int_\M L$, it is natural to regard $\SS_L$ as the physically relevant object. The converse, ``$\SS_L=\SS_{L'}$ implies $L'=L+\d Y$'', holds under suitable topological conditions on $\F$ \cite{margalef2021geometric} that we assume hereafter. Once again, the Lagrangian is ambiguous, but its cohomology class is not. In particular, any physically meaningful derived quantity must vanish for the trivial Lagrangian $L=\d Y$, since $\SS_{\d Y}=0$. For instance, \cref{eq: first variation} with $L=\d Y$ reduces to
\begin{equation}\label{eq: first variation L=dY}
    \dd L=\dd(\d Y)=\d \dd Y
\end{equation}
so $\mathrm{EOM}=0$ (also obvious using $\SS_{\d Y}=0$), $\Theta=\dd Y+\d Z$ (two sources of ambiguity), and $[\Theta]=[\dd Y]$ (one source of ambiguity). Thus, EOM is a physical quantity but $\Theta$ and $[\Theta]$ are not. The symplectic current is also non-physical since for $L=\d Y$ we have
\[\omega=\dd(\dd Y+\d Z)=\dd^2 Y+\d \dd Z=\d(\dd Z)\]
However, the (pre)symplectic form is independent of $(Y,Z)$ as, upon integration via Stokes' theorem, it yields the trivial presymplectic form $\Omega_\Sigma=0$.

Finally, the choice of Cauchy surface $\Sigma$ affects the presymplectic form over $\F$. Indeed, given two disjoint Cauchy surfaces $\Sigma_1, \Sigma_2$ bounding a region $\M_{12}$, using Stokes' theorem, we have 
\[\Omega_{\Sigma_2}-\Omega_{\Sigma_1}=\int_{\M_{12}}\d\omega=\int_{\M_{12}}\dd\d\Theta=-\int_{\M_{12}}\dd\mathrm{EOM}_i\wwedge\dd\Phi^i\]
Although $\Omega$ depends on the chosen Cauchy surface, its pull-back to the space of solutions (called on-shell) does not. Indeed, if $\jj:\Sol(\SS)\hookrightarrow\F$ is the inclusion into the solution space, then $\jj^*\Omega_{\Sigma_2}=\jj^*\Omega_{\Sigma_1}$ since $\jj^*\mathrm{EOM}\equiv 0$. This (pre)symplectic form, denoted $\Omega_\SS$, is canonically associated with $\SS$.

%/=-=-=-=-=-=-=\%
\subsection{Summary of the variations}
%\=-=-=-=-=-=-=/%
For any local action $\SS$, the CPS formalism provides, a unique (pre)symplectic space of solutions:
\[\Psi:\Omega^{0}(\F)\to\mathrm{Sympl}\qquad\text{given by}\qquad\Psi(\SS)=\Big(\{\Phi\in\F:\mathrm{EOM}_i(\Phi)=0\},\Omega_\SS\Big)\] 
defined via \cref{eq: first variation,eq: omega=dlTheta,eq: Omega=int omega} using any representative $L$ of $\SS$. In practice, it is better to operate at the Lagrangian level, where the relevant map is
\begin{equation}\label{eq: Lambda L first variation}
    \begin{array}{cccc}
\Lambda:&\Omega^{(n,0)}(\M\times\F)&\longrightarrow&\Omega^{n}(\M)\times\Omega^{(n-1,1)}(\M\times\F)\\
&L&\longmapsto&(\mathrm{EOM}_i,\Theta+\d Z)\end{array}
\end{equation} 
The symplectic potential is determined up to the $\d Z$ ambiguity discussed in \cref{subsection: Ambiguities}.

\begin{remarks}
    \item $\Psi(\SS_{\d Y})=(\F,0)$. The trivial action yields the full field space with a zero presymplectic form.
    \item  $\Lambda(\d Y)=(0,\dd Y+\d Z)$. A total derivative Lagrangian has vanishing equations of motion, and its symplectic potential inherits the aforementioned ambiguities.  
\end{remarks}

%/=-=-=-=-=-=-=\%
\subsection{Variational vector fields}
%\=-=-=-=-=-=-=/%
In classical mechanics, transformations like rotations or translations take place in the space of coordinates. Rather than working with finite transformations, which form a group, we usually work with infinitesimal transformations, which form an algebra, i.e., vector fields over the space of coordinates. In field theories, the role of coordinates is played (loosely speaking) by tensor fields, so an infinitesimal transformation is a vector field $\XX\in\mathfrak{X}(\F)$, called a variational vector field (VVF).

For a tensorial field $\Phi^i$, we define $\XX_{\Phi^i}:=\Psi^i\in T_{\Phi}\F$. Here, $\Psi^i$ satisfies the linearized conditions of $\Phi^i$ so, in particular, shares its index character. Setting aside infinite-dimensional subtleties \cite{margalef2021geometric}, $\XX$ can be equivalently understood as a derivation via the Lie derivative of $\F$:
\begin{equation}\label{eq: definition XX}
\XX_{\Phi^i}:=\Psi^i\qquad\qquad\equiv\qquad\qquad\LL_{\XX}\Phi^i:=\Psi^i\qquad\qquad\equiv\qquad\qquad\ii_{\XX}\dd\Phi^i:=\Psi^i
\end{equation}
\begin{remark}
    The last two formulas are related via Cartan's magic formula over $\F$ (which holds for a broad class of VVFs but not all, see \cref{subsection: Vertical operators}): $\LL_{\XX}=\ii_{\XX}\dd+\dd\ii_{\XX}$. 
\end{remark}
The last expression of \cref{eq: definition XX} serves as an implicit definition of $\XX$ as a map acting on $1$-forms via the interior product. This enables a formal expression analogous to the finite-dimensional case:
\begin{equation}\label{eq: XX=int delta delta}
\xi=\sum_{j=1}^n\iota_\xi(\d x_j)\frac{\partial}{\partial x_j}\in\mathfrak{X}(\M)\qquad\qquad\qquad
    \XX=\int_\M\ii_\XX( \dd\Phi^i)\frac{\delta}{\delta \Phi^i}\in\mathfrak{X}(\F)
\end{equation}
These expressions conveniently package their respective components, $\{\iota_\xi(\d x_j)\}_j$ and $\{\ii_\XX (\dd\Phi^i)\}_i$, into a single geometric object. While  \cref{eq: XX=int delta delta} is purely formal, it is highly practical for implementing VVFs in \xCPS, as demonstrated in \cref{section: VVF}.

\begin{example}{}{}
Given $\xi\in\mathfrak{X}(\M)$, we define $\XX_\xi\in\mathfrak{X}(\F)$ as the infinitesimal action of $\mathrm{Diff}(\M)$:
\begin{equation}\label{eq: X_xi}
    \ii_{\XX_\xi} \dd\Phi^i:=\L_\xi\Phi^i\quad\quad\equiv\quad\quad\LL_{\XX_\xi}\Phi^i:=\L_\xi\Phi^i\quad\quad\equiv\quad\quad\XX_\xi=\int_\M\L_\xi\Phi^i\frac{\delta}{\delta \Phi^i}
\end{equation}
The action of $\XX_\xi$ generates, loosely speaking, the transformation $\Phi^i\mapsto\Phi^i+\varepsilon\L_\xi\Phi^i+\mathcal{O}(\varepsilon^2)$. Notice that in the physics literature, this is frequently denoted as $\delta_\xi\Phi=\L_\xi\Phi^i$.
\end{example}

%/=-=-=-=-=-=-=\%
\subsection{Noether symmetries and charges}
%\=-=-=-=-=-=-=/%
\begin{definition}{}{}
Given a local action $\SS$, we say that a VVF, $\XX$, is a symmetry of $\SS$ if $\LL_{\XX}\SS=0$.
\end{definition}
Such a symmetry leaves the action $\SS$ (and therefore the physics) invariant, which subsequently preserves the set of critical points $\Sol(\SS)$. Thus, its restriction $\overline{\XX}:=\XX|_\Sol(\SS)$ is a vector field of $\Sol(\SS)$. A natural question is whether $\overline{\XX}$ is a Hamiltonian vector field with respect to $\Omega_\SS$. Among all symmetries, the $\d$-symmetries defined below admit a systematic treatment via Noether's theorem.
\begin{definition}{}{}
A VVF, $\XX$, is a $\d$-symmetry (or variational symmetry) of $L$ if
\begin{equation}\label{eq: L_X L=d S}
\LL_{\XX}L=\d S_\XX
\end{equation}
for some $S_\XX\in\Omega^{(n-1,0)}(\M\times\F)$. In this case, $S_\XX$ is called a symmetry potential of $\XX$.
\end{definition}

\begin{remarks}
    \item If $\XX$ is a $\d$-symmetry of $L$, then it is also a $\d$-symmetry of any $L'=L+\d Y$.
    \item If $\XX$ is a $\d$-symmetry of $L$, then $\XX$ is a symmetry of $\SS_L$. The converse is not generally true.
    \item $S_\XX$ is defined up to a closed form, which might not be exact. For simplicity, we assume that every closed $(n-1)$-form is exact (see \cite{margalef2021geometric} for a discussion when this does not hold).
\end{remarks}

Applying Cartan's magic formula and \cref{eq: first variation}, we find:
\[0=\LL_{\XX}L-\d S_\XX=\ii_{\XX}\dd L-\d S_\XX=\ii_{\XX}(\mathrm{EOM}_i\dd\Phi^i+\d\Theta)-\d S_\XX=\mathrm{EOM}_i\LL_{\XX}\Phi^i-\d(S_\XX-\ii_\XX\Theta)\]

\begin{definition}{}{}
Given a $\d$-symmetry $\XX$ of $\SS_L$, we define its $\XX$-current (or Noether current) as
\[J_\XX=S_\XX-\ii_\XX\Theta\in\Omega^{(n-1,0)}(\M\times\F)\]
\end{definition}
Integrating over a Cauchy surface $\Sigma$ yields the $\XX$-charge (or Noether charge):
\begin{equation}
    \HH_\XX:=\int_\Sigma \imath^*J_\XX\qquad\qquad\longrightarrow\qquad\qquad\HH_{\overline{\XX}}:=\jj^*\HH_\XX
\end{equation}
Tracking the ambiguities outlined in \cref{subsection: Ambiguities}, one finds that $\HH_\XX$ depends only on $\SS$, $\XX$, and $\imath:\Sigma\hookrightarrow\M$, while its pull-back to the space of solutions $\HH_{\overline{\XX}}$  (on-shell), depends exclusively on $\SS$ and $\overline{\XX}$. This latter quantity represents the physically meaningful conserved Noether charge, taking the same value on any Cauchy surface $\Sigma$. Furthermore, $\HH_{\overline{\XX}}$ is the Hamiltonian of $\overline{\XX}$, i.e., $\ii_{\overline{\XX}}\Omega_\SS=\dd\HH_{\overline{\XX}}$. Finally, we define a gauge transformation as a VVF over the space of solutions $\overline{\XX}$ such that:
\begin{equation}\label{eq: gaugue transformation}
    \ii_{\overline{\XX}}\Omega_\SS=0\qquad\quad\equiv\quad\qquad \HH_{\overline{\XX}}=0 \text{ (up to a constant)}\qquad\quad\equiv\qquad\quad J_{\overline{\XX}}=\d Q_{\overline{\XX}}
\end{equation}
\begin{example}{}{ first variation EM}
    Let us consider the electromagnetic (EM) Lagrangian given by
    \[L_{\mathrm{EM}}(A)=\frac{1}{4}g^{ac}g^{bd}(\d A)_{ab}(\d A)_{cd}\vol_g\]
    where $A\in\Omega^1(\M)$ and $g$ is some fixed metric. A straightforward computation shows that
    \[\mathrm{EOM}_{\mathrm{EM}}^b=-\nabla_a(\d A)^{ab}\vol_g\qquad\qquad \Theta_{\mathrm{EM}}=\iota_{\vec{U}}\vol_g\quad\text{with}\quad U^a=(\d A)^{ab}(\dd A)_b\]
    Letting $\lambda\in\Omega^0(\M)$, we define $\XX_\lambda$ by:
\begin{equation}\label{eq: X_lambda}
\ii_{\XX_\lambda} \dd A:=\d\lambda\quad\qquad\equiv\quad\qquad\LL_{\XX_\lambda}A:=\d\lambda\qquad\quad\equiv\quad\qquad\XX_\lambda=\int_\M\d\lambda\frac{\delta}{\delta A}\end{equation}
Since $\d^2=0$, it is easy to check that
\[\LL_{\XX_\lambda}L_{\mathrm{EM}}=0\quad\qquad\longrightarrow\quad\qquad J_{\XX_\lambda}=-\ii_{\XX_\lambda}\Theta=-\iota(\ii_{\XX_\lambda}\vec{U})\vol_g=-\iota((\d A)^{ab}(\d\lambda)_b)\vol_g\]
Integrating by parts produces the Noether charge:
\[\HH_{\XX_\lambda}=\int_\Sigma\lambda n_a \mathrm{EOM}_{\mathrm{EM}}^a\vol_\gamma\qquad\qquad\longrightarrow\qquad\qquad\HH_{\overline{\XX}_\lambda}=0\]
Thus, $\XX_\lambda$ is a $\d$-symmetry and $\overline{\XX}_\lambda$ is, according to \cref{eq: gaugue transformation}, a gauge transformation.
\end{example}

%/=-=-=-=-=-=-=\%
\subsection{Diff-invariance and non-Noether charges}\label{subsection: Non-Noether}
%\=-=-=-=-=-=-=/%
An action $\SS$ is Diff-invariant if $\XX_{\xi}$, as defined in \cref{eq: X_xi}, is a symmetry of $\SS$ for all $\xi\in\mathfrak{X}(\M)$. A Lagrangian $L$ is Diff-invariant if $\LL_{\XX_{\xi}}L=\L_\xi L$ for all $\xi\in\mathfrak{X}(\M)$. If $L$ is Diff-invariant, then $\XX_{\xi}$ is a $\d$-symmetry of $L$ (though the converse might not hold) with
\begin{equation}\label{eq: Diff invariance XX_xi}
    S_{\XX_\xi}=\iota_\xi L\quad\longrightarrow\quad J_{\XX_\xi}=\iota_\xi L-\ii_{\XX_\xi}\Theta\quad\longrightarrow\quad\HH_{\XX_\xi}:=\int_\Sigma \imath^*J_{\XX_\xi}\quad\longrightarrow\quad\HH_{\overline{\XX}_\xi}:=\jj^*\HH_{\XX_\xi}
\end{equation}
Notably, these quantities can be defined even when $\XX_\xi$ is not a $\d$-symmetry:
\begin{equation}
    S_\xi:=\iota_\xi L \quad\longrightarrow\quad J_\xi:=\iota_\xi L-\ii_{\XX_\xi}\Theta \quad\longrightarrow\quad \QQ^\Sigma_\xi:=\int_\Sigma \imath^*J_\xi\quad\longrightarrow\quad\overline{\QQ}^\Sigma_\xi:=\jj^*\QQ_{\xi}^\Sigma
\end{equation}
that we call $\xi$-potential, $\xi$-current, off-shell $\xi$-charge, and on-shell $\xi$-charge. If $\XX_\xi$ is not a symmetry, $\overline{\QQ}^\Sigma_\xi$ is a non-Noether charge. This generally depends on the chosen Lagrangian representative and the Cauchy surface $\Sigma$, leading to a flux law relating charges on different Cauchy surfaces.
\begin{example}{}{}
The $\xi$-charge of the Lagrangian $L_\mathrm{scl}$ given by \cref{eq: Lagrangian scalar field} is:
    \[\QQ_\xi^\Sigma=\int_\Sigma n_a\xi_bT^{ab}_{\mathrm{scl}}\vol_\gamma\]
    Here, $T^{ab}_{\mathrm{scl}}=\nabla^a\phi\nabla^b\phi-g^{ab}L_{\mathrm{scl}}(\phi)$ is the energy-momentum tensor. 
\end{example}

%/=-=-=-=-=-=-=\%
\subsection{Top-forms vs densities}\label{subsection: top-forms vs densities}
%\=-=-=-=-=-=-=/%
Because \xAct operates on tensorial expressions rather than geometric integrals, \xCPS works with densitized currents instead of top-forms and their integrated counterparts. These approaches are equivalent once a fiducial volume form $\vol_0$ on $\M$ is fixed, which we assume henceforth. We now explicitly define $\widetilde{\Lambda}$, the density-level analogue to $\Lambda$ from \cref{eq: Lambda L first variation}.

First consider a Lagrangian function $\mathcal{L}\in\Omega^{(0,0)}(\M\times\F)$. Clearly $L:=\mathcal{L}\vol_0\in\Omega^{(n,0)}(\M\times\F)$ is a Lagrangian with the usual decomposition $\dd L=\mathrm{EOM}\dd\Phi^i+\d\Theta$ given by \cref{eq: first variation}. Since $\Theta$ has horizontal-co-degree $1$, there exists a symplectic potential vector $\vec{U}$ such that $\Theta=\iota_{\vec{U}}\vol_0$. Therefore, the same computation used after \cref{eq: first variation scalar field 2} shows that $\d\Theta=\mathrm{div}(\vec{U})\vol_0$. Thus, using the fact that $\vol_0$ is fixed ($\dd \vol_0=0)$, we obtain the density-equivalent decomposition to \cref{eq: first variation}:
\begin{equation}\label{eq: first variation densities}
  \dd \mathcal{L}=\frac{\dd L}{\vol_0}=\frac{\mathrm{EOM}_i}{\vol_0}\dd\Phi^i+\frac{\d\Theta}{\vol_0}=\frac{\mathrm{EOM}_i}{\vol_0}\dd\Phi^i+\frac{\mathrm{div}(\vec{U})\vol_0}{\vol_0}=\mathrm{eom}_i\dd\Phi^i+\mathrm{div}(\vec{U})
\end{equation}
Here, $\mathrm{eom}_i$ is the densitized equations of motion (i.e. $\mathrm{EOM}_i=\mathrm{eom}_i\vol_0$), whose kernel is the same as $\mathrm{EOM}_i$. Consequently, we define the canonical decomposition map $\widetilde{\Lambda}$ as:
\[\begin{array}{cccc}
\widetilde{\Lambda}:&\Omega^{(0,0)}(\M\times\F)&\longrightarrow&\Omega^{(0,0)}(\M\times\F)\times\mathcal{T}^{(1,1)}(\M\times\F)\\
&\mathcal{L}&\longmapsto&(\ \mathrm{eom}_i\ ,\ \vec{U}\ )\end{array}\] 
Notice that $\vec{U}$ is a vector in $\M$ and a $1$-form in $\F$ (see \cref{example: first variation scalar,example: first variation EM}).
\begin{remarks}
    \item $\widetilde{\Lambda}(\mathrm{div}\vec{V})=(0,\mathrm{div}(\dd\vec{V}))$ with an additional ambiguity similar to $\d Z$.
    \item Upon integration over a Cauchy surface, we have:
    \begin{equation}\label{eq: Omega=n_a U^a vol Sigma}
    \Omega =\int_\Sigma\imath^*\dd\Theta=\int_\Sigma\imath^*\iota_{\dd \vec{U}}\vol_0=-\int_\Sigma n_a\dd U^a\vol_\Sigma
    \end{equation}
    where $n_a$ is a $1$-form transverse to $\Sigma$ (i.e. $\vol_0=-n\wedge \vol_\Sigma$ over $\Sigma$).
\end{remarks}

Likewise, we establish a density equivalent to \cref{eq: L_X L=d S}. Because $S_\XX$ has a horizontal co-degree of $1$, there exists a symmetry potential vector $\vec{s}_\XX$ such that $S_\XX=\iota_{\vec{s}_\XX}\vol_0$. This gives:
\begin{equation}\label{eq: L_XL=div s}
\LL_{\XX}\mathcal{L}=\frac{\LL_{\XX} L}{\vol_0}=\frac{\d S_\XX}{\vol_0}=\mathrm{div}(\vec{s}_\XX)
\end{equation}

Finally, the densitized Noether current and non-Noether current are defined as:
\begin{equation}\label{eq: Noether current density}
    \vec{j}_\XX=\vec{s}_\XX-\ii_\XX\vec{U}\qquad\qquad\vec{j}_\xi=\mathcal{L}\vec{\xi} -\ii_{\XX_\xi}\vec{U}
\end{equation}

% dl(div(V))=dl(div(V)vol)/vol=dl(L_Vvol)/vol=dl(di_Vvol)/vol=d(dli_Vvol)/vol=di_{dlV}vol/vol=L_{dlV}vol/vol=div(dlV)vol/vol=div(dlV)
%\ii_\XX\vec{U}\qquad\qquad\imath^*j_\xi=\iota_\xi L-\ii_{\XX_\xi}\Theta=\iota_\xi (-\mathcal{L}n\wedge\vol_\Sigma)+\ii_{\XX_\xi}\iota_{\vec{U}}(n\wedge\vol_\Sigma)=-n_a(\mathcal{L}\xi^a -\ii_{\XX_\xi}\vec{U}^a)
% \vec{s}_{j_\xi}\vol=S_\xi=\iota_{\xi}L=\iota_{\xi \mathcal{L}}\vol
\begin{remark}
    Several signs appearing in the equations of this section are conventional and can be globally adjusted within the \xCPS package settings.
\end{remark}

%%%%%%%%%%%%%%%%%%%%%%%%%%%%%%%%%%%%%%%%%%%%%%%%%%%%%%%%%%%%%%%%%%%%%%%%%%%%%%%%%%%%%%%%%%
%%%%%%%%%%%%%%%%%%%%%%%%%%%%%%%%%%%%%%%%%%%%%%%%%%%%%%%%%%%%%%%%%%%%%%%%%%%%%%%%%%%%%%%%%%
%%%%%%%%%%%%%%%%%%%%%%%%%%%%%%%%%%%%%%%%%%%%%%%%%%%%%%%%%%%%%%%%%%%%%%%%%%%%%%%%%%%%%%%%%%

%/=====================\%
\section{Computational background}\label{sec: xact}
%\=====================/%
%/=-=-=-=-=-=-=\%
\subsection{xAct}
%\=-=-=-=-=-=-=/%
\xAct is a \mathematica bundle for symbolic tensor algebra. Its core functionality is distributed across several packages: \xPerm handles index canonicalization under permutation symmetries \cite{xPerm}; \xCore provides basic programming tools; \xTensor extends \mathematica with abstract tensor calculus; and \xCoba complements \xTensor with  tools for explicit bases and component calculations.

In this section, we provide a brief introduction to \xTensor. For brevity, we suppress all information messages (see \cite{xCPS} for the companion file); for clarity, outputs are slightly formatted when needed (\LaTeX{} formulas are retrieved via \TexAct \cite{TexAct}). We load the core framework with the command:
\setcounter{mmaCounter}{1}
\begin{mmaIn}
<{}<xAct\`{}xTensor\`{}
\end{mmaIn} 

%/=-=-=-=-=-=-=\%
\subsection{Quick tour of xAct}\label{subsection: quick tour xact}
%\=-=-=-=-=-=-=/%
The most basic command is \x{DefManifold}, which defines a manifold and its tangent bundle. Following Penrose's abstract index notation, a set of abstract indices must be explicitly declared. The dimension can be a fixed integer or a symbolic variable.
\begin{mmaIn}
DefManifold[M,4,\{a,b,c,d,e,f\}]
\end{mmaIn} 
Once a manifold is defined, tensors are declared with \x{DefTensor}, requiring the user to specify index slots as covariant or contravariant within previously defined bundles. Monoterm symmetries can be imposed directly on the indices. However, there are no known algorithms to handle multiterm symmetries (such as Bianchi identities), though these can sometimes be implemented manually.
\begin{mmaIn}
DefTensor[T[a,b,-c],M,Symmetric[\{a,b\}]]
\end{mmaIn} 
Since $T$ is symmetric in its first two indices, the expression $\tensor{T}{^a^b_c}-\tensor{T}{^b^a_c}$ should be zero. However, \mathematica would not simplify this expression even if we use \x{Simplify}, because it is not aware of this symmetry. \xAct has that information and can rewrite this expression using \x{ToCanonical}:

\begin{mmaIn}
T[a,b,-c]-T[b,a,-c] // Simplify (* Symmetry not considered *)\newline
T[a,b,-c]-T[b,a,-c] // ToCanonical
\end{mmaIn} 
\begin{mmaOut}
$\tensor{T}{^a^b_c}-\tensor{T}{^b^a_c}$ \\
0
\end{mmaOut} 
\x{ToCanonical} is arguably the most important command in \xAct, as it brings any expression down to a unique canonical form. While canonicalization does not necessarily yield a simpler-looking expression, it enforces a standard form which is suited for \mathematica's simplification algorithms. For convenience, \xAct provides the command \x{Simplification} which is \x{Simplify@ToCanonical}.

With a manifold and tensors defined, we can perform differential calculus. Without introducing additional structures, we have access to Lie derivatives, \x{LieD}, and Lie Brackets, \x{Bracket}.
\begin{mmaIn}
DefTensor[\{v[a],w[a]\},M]\mmaSup\newline
LieD[v[d]]@T[a,b,-c]\newline
Bracket[v[d],w[d]][c]
\end{mmaIn}
\begin{mmaOut}
$\mathcal{L}_v\tensor{T}{^b^a_c}$ \\
$[v^d,w^d]^c$
\end{mmaOut} 
The vectors \x{v[d]} and \x{w[d]} are the arguments and, when playing this operational role, their indices are called ``ultraindices''. They act purely as placeholders identifying the vectors and do not appear in the index structure of the output, and may be freely repeated.

The previous expressions can be rewritten in terms of some covariant derivatives (CovD). By default, \xAct uses a fiducial torsion-free flat connection named \x{PD}, denoted $\partial$, and referred to as the ordinary derivative in \cite{wald2010general}. It is not unique yet constitutes a CovD in its own right (no coordinates needed!).
\begin{mmaIn}
LieDToCovD[LieD[v[d]]@T[a,b,-c]]\newline
BracketToCovD[Bracket[v[d],w[d]][c]]
\end{mmaIn}
\begin{mmaOut}
$ v^d  \partial_{d}\tensor{T}{^a^b_c}+\tensor{T}{^a^b_d}\partial_cv^d -\tensor{T}{^d^b_c}\partial_dv^a-\tensor{T}{^a^d_c}\partial_dv^b$ \\
$v^a  \partial_{a}
w^c- w^a\partial_{a}v^c $
\end{mmaOut} 
Additional CovDs, possibly with torsion, can be defined on any tangent bundle:
\begin{mmaIn}
DefCovD[CD[-a],\{";","D"\},Torsion$\to$True]
\end{mmaIn}
This defines the covariant derivative \x{CD} over \x{M} (to which the index \x{a} belongs), denoted \x{D}, along with its associated tensors: \x{TorsionCD[a,-b,-c]} (zero if \x{Torsion$\to$False}), \x{ChristoffelCD[a,-b,-c]}, \x{RiemannCD[-a,-b,-c,d]} and \x{RicciCD[-a,-b]}. Because this CovD is non-metric, \x{ChristoffelCD} and \x{RicciCD} are not symmetric and \x{RiemannCD} is only antisymmetric in the first pair.
\begin{remark}
    \x{ChristoffelCD} is a genuine tensor, not merely a connection symbol. It measures the difference between \x{CD} and the fiducial connection \x{PD}:
    
\begin{mmaIn}
CD[-a]@T[b,-c,-d] == (CD[-a]@T[b,-c,-d] // ChangeCovD)
\end{mmaIn}
\begin{mmaOut}
$D_{a}T^{b}{}_{cd} =\partial_{a}T^{b}{}_{cd} + \Gamma [D]^{b}{}_{ae} T^{e}{}_{cd} - \Gamma [D]^{e}{}_{ad} T^{b}{}_{ce} -  \Gamma [D]^{e}{}_{ac} T^{b}{}_{ed} $
\end{mmaOut}
\end{remark}

We can also define inner vector bundles (possibly complex) over \x{M}, equipped with their corresponding abstract indices, where the specified dimension corresponds to the dimension of a typical fiber.
\begin{mmaIn}
DefVBundle[inner,M,3,\{J,P,Q,X\}]
\end{mmaIn}
Subsequently, we can define a CovD that acts simultaneously on both \x{TangentM} and \x{inner} indices.
\begin{mmaIn}
DefCovD[CD2[-a],inner,\{";","$\nabla$"\},Torsion$\to$True]
\end{mmaIn}
In addition to the standard tensors associated with any CovD (\x{TorsionCD2}, \x{ChristoffelCD2}, \x{RiemannCD2}, and \x{RicciCD2}), two bundle-specific tensors are defined: \x{AChristoffelCD2[J,-b,-P]}, which measures the difference between \x{PD} and  \x{CD2} for inner indices (\x{b} is the derivative index), and \x{FRiemannCD2[-a,-b,-P,Q]}, the curvature tensor on the inner bundle.

\begin{mmaIn}
DefTensor[$\beta$[-a,J],M]\mmaSup\newline
CD2[-a]@$\beta$[-b,J] == (CD2[-a]@$\beta$[-b,J] // ChangeCovD)
\end{mmaIn}
\begin{mmaOut}
$\nabla_{a}\beta_{b}{}^{J} =\partial_{a}\beta_{b}{}^{J}-  \Gamma [\nabla ]^{c}{}_{ab} \beta_{c}{}^{J} +  A[\nabla ]^{J}{}_{aP} \beta_{b}{}^{P} $
\end{mmaOut}

Mathematically, the CovD of a tensor is another tensor. Computationally, however, \xAct treats derived tensors differently: they are handled as nested operations rather than formally declared tensor objects. As such, they are not stored in \x{\$Tensors}, which catalogs ``properly defined'' tensors (those created via \x{DefTensor}, \x{DefCovD}, etc.).
\begin{mmaIn}%
\$Tensors
\end{mmaIn}
\begin{mmaOut}
\{T,v,w,TorsionCD,ChristoffelCD,RiemannCD,RicciCD,TorsionCD2,ChristoffelCD2,\newline RiemannCD2,RicciCD2,AChristoffelCD2,FRiemannCD2,$\beta$\}
\end{mmaOut}
When studying higher-order derivatives (as detailed in \cref{sec: final examples}), it is necessary to treat these derived tensors on equal footing with bare tensors. For that, \xAct includes the command \x{Implode}, which ``implodes'' (merges) the CovD symbol with the tensor symbol. The first time it is used, a new tensor is declared via \x{DefTensor}. For instance:
\begin{mmaIn}
Implode[CD[-d]@T[a,b,-c]]\newline
Implode[CD[-e]@CD[-d]@T[a,b,-c]]
\end{mmaIn}
\begin{mmaOut}
$\tensor{DT}{^a^b_c_d}$\\
$\tensor{DDT}{^a^b_c_d_e}$
\end{mmaOut}
These commands define the imploded tensors \x{CDT} and \x{CDCDT}, given by $\tensor{(D T)}{^a^b_c_d}:=D_d \tensor{T}{^a^b_c}$ and $\tensor{(D\!DT)}{^a^b_c_d_e}:=D_eD_d \tensor{T}{^a^b_c}$ (derivative indices at the end), and which are now stored on \x{\$Tensors}:
\begin{mmaIn}
\$Tensors
\end{mmaIn}
\begin{mmaOut}
\{T,v,w,TorsionCD,ChristoffelCD,RiemannCD,RicciCD,TorsionCD2,ChristoffelCD2, RiemannCD2,RicciCD2,AChristoffelCD2,FRiemannCD2,$\beta$,CDT,CDCDT\}
\end{mmaOut}
Imploded tensors can be written in exploded form (split) using the command \x{Explode}:
\begin{mmaIn}
Explode[CDT[a,b,-c,-d]]
\end{mmaIn}
\begin{mmaOut}
$D_{d}\tensor{T}{^a^b_c}$
\end{mmaOut}

One final important command is \x{DefMetric}, which defines a metric of a given signature $\pm 1$ on \x{M} and its associated Levi-Civita connection. As with any CovD, \x{DefMetric} defines \x{Christoffel}, \x{Riemann}, and \x{Ricci}. The metric provides additional curvature tensors like, among others, the Ricci scalar \x{RicciScalar}, the metric volume form \x{epsilong}, and the determinant density \x{Detg}.

Once a metric is defined, indices of \x{M} can be raised and lowered using \x{ContractMetric}.
\begin{mmaIn}
DefMetric[-1,g[-a,-b],LCDer,\{";","$\mathcal{D}$"\}]\mmaSup\newline
\{v[a]g[-a,-b],T[a,b,-c]g[-a,-b],g[-b,-c]LCDer[-a]@v[c]\}\newline
\{v[a]g[-a,-b],T[a,b,-c]g[-a,-b],g[-b,-c]LCDer[-a]@v[c]\} // ContractMetric
\end{mmaIn}
\begin{mmaOut}
$\{g_{ab}v^a, g_{ab}\tensor{T}{^a^b_c}, g_{bc}\mathcal{D}_av^c\}$\\
$\{v_b, \tensor{T}{^a_a_c}, \mathcal{D}_av_b\}$
\end{mmaOut}
 
%%%%%%%%%%%%%%%%%%%%%%%%%%%%%%%%%%%%%%%%%%%%%%%%%%%%%%%%%%%%%%%%%%%%%%%%%%%%%%%%%%%%%%%%%%
%%%%%%%%%%%%%%%%%%%%%%%%%%%%%%%%%%%%%%%%%%%%%%%%%%%%%%%%%%%%%%%%%%%%%%%%%%%%%%%%%%%%%%%%%%
%%%%%%%%%%%%%%%%%%%%%%%%%%%%%%%%%%%%%%%%%%%%%%%%%%%%%%%%%%%%%%%%%%%%%%%%%%%%%%%%%%%%%%%%%% 

%/=====================\%
\section{xCPS core functionalities}\label{sec: functionalities}
%\=====================/%
This section explains the core functionalities of \xCPS, which is distributed as part of \xAct under the GNU General Public License \cite{xCPS} and is also available at \url{https://github.com/juanmargalef/xCPS}. Where possible, explanations are language-agnostic to facilitate reimplementation in other languages. We begin a new \mathematica session and load \xCPS (which automatically loads \xAct) using:
\setcounter{mmaCounter}{1}
\begin{mmaIn}
<{}<xAct\`{}xCPS\`{}\newline
SetOptions[ContractMetric,AllowUpperDerivatives$\to$True];\newline
DefManifold[M,4,\{a,b,c,d,e,f,h,i,j,k,l\}]
\end{mmaIn}

%/=-=-=-=-=-=-=\%
\subsection{Vertical forms and the field-space wedge product}
%\=-=-=-=-=-=-=/%
\x{DefTensor} has been modified in three key ways: 

\begin{enumerate}
    \item It accepts the additional option \x{VertDeg}, a non-negative integer (0 by default) indicating that the defined tensor of \x{M} is also a form on the space of fields $\F$ of that degree. The term ``vertical'' refers to the fiber direction in the variational bicomplex \cite{margalef2021geometric,anderson1989variational} as opposed to the ``horizontal'' degree on \x{M}.
    \item For any tensor \x{T}, an additional tensor \x{dlT} is defined: the vertical exterior derivative of \x{T}, printed as $\dd\x{T}$, with the same characteristics as \x{T} except that \x{VertDeg[$\dd$T]=VertDeg[T]+1}.\footnote{The prepend ``\x{dl}'' and the print symbol $\dd$ can be modified via \x{\$NameVerticalExteriorDerivative} and \x{\$SymbolVerticalExteriorDerivative}.} 
    \item For any tensor \x{T}, an additional tensor \x{VariationalVectorT} is defined: the dual to the first-order variation of \x{T}, printed as $\frac{\delta}{\delta\x{T}}$. It has ``opposite'' characteristics to \x{T}: opposite variance of indices (but identical symmetries), opposite vertical degree, \x{VertDeg[$\frac{\delta}{\delta\x{T}}$]=-VertDeg[T]}, and opposite density weight, \x{WeightOfTensor[$\frac{\delta}{\delta\x{T}}$]=-WeightOfTensor[T]}.
\end{enumerate}
\begin{remark}
$\dd\x{T}$ and $\frac{\delta}{\delta\x{T}}$ are dual: their contraction yields a scalar (see details on \cref{section: ExpandVertInt}).
\end{remark}
 
The vertical degree enables a new bilinear supergraded product called \x{WWedge}, corresponding to the natural wedge of $k$-forms on the space of fields:
\begin{equation}
    \alpha\wwedge\beta=(-1)^{\x{VertDeg[$\alpha$]}\x{VertDeg[$\beta$]}}\beta\wwedge\alpha
\end{equation}

\begin{remarks}
    \item The \x{WWedge} product is associative and reduces to the standard \x{Times} product if at least one of the two terms has a \x{VertDeg} of zero.
    \item Several \xAct commands have been modified to account for the \x{WWedge} product. For instance, \x{LieD} and \x{Bracket} must satisfy the Leibniz rule with respect to the \x{WWedge}.
    \item Part of the setup so far is the vertical analog of the \xTerior package \cite{xTerior} (which implements the horizontal \x{Wedge} product of horizontal forms). A key difference is that horizontal $k$-forms vanish when $k$ exceeds the dimension of \x{M}, while there is no limit to the degree of the vertical differential forms ($\F$ is infinite-dimensional).
\end{remarks}

%/=-=-=-=-=-=-=\%
\subsection{Variational vectors and variational vector fields}\label{section: VVF}
%\=-=-=-=-=-=-=/%
As mentioned before, each tensor defined via \x{DefTensor} has an associated variational vector with ``opposite'' characteristics. For example, for a metric $g_{ab}$, its variational vector $\frac{\delta}{\delta g}$ has two upper symmetric indices:
\begin{equation}\label{eq: delta delta example}
\left(\frac{\delta}{\delta g}\right)^{ab}
\end{equation}

 \begin{remarks}
     \item While physics literature frequently uses the notation $\frac{\delta}{\delta g_{ab}}$, for implementation purposes, it is best to consider the notation of \cref{eq: delta delta example}.
     \item In certain physical contexts, it is preferable to consider variations of the inverse metric $(g^{-1})^{ab}$. \xCPS supports this by setting \x{\$UseInverseMetric=True} (\x{False} by default), which automatically applies the replacement rules:
    \begin{equation}
        \left(\frac{\delta}{\delta g}\right)^{ab}=-g^{ac}g^{bd}\left(\frac{\delta}{\delta g^{-1}}\right)_{cd}\qquad\qquad (\dd g)_{ab}=-g_{ac}g_{bd}\dd (g^{-1})^{cd}
    \end{equation}
\end{remarks}
Variational vector fields (VVFs) are then defined in terms of variational vectors as in  \cref{eq: XX=int delta delta} (see also \cref{eq: X_lambda,eq: X_xi}). Concretely, a VVF is a scalar expression (no free indices) in which each monomial contains exactly one variational vector, such as
\begin{equation}\label{eq: VVF example}
    \tensor{T}{^a_b_c}\tensor{\left(\frac{\delta}{\delta S}\right)}{_a^b^c}+R^{ab}\wwedge\tensor{\left(\frac{\delta}{\delta Q}\right)}{_a_b}
\end{equation}
The formal integral of \cref{eq: XX=int delta delta} is omitted: \xAct handles the expression directly, and including the integral would add no computational value. What matters  computationally is how a VVF acts on differential forms, which is the subject of \cref{subsection: Vertical operators}.

\xCPS provides a streamlined command to define the VVF $\XX_\xi$ (see \cref{eq: X_xi}) for a given vector field $\xi^a$ of \x{M}: \x{VVFFromLieD[vector][tensors]}. It returns the variational vector field
\begin{equation}
    \sum_{\x{tensor}\in\{\x{tensors}\}}\x{LieD[vector][tensor]}\wwedge \x{VariationalVector[tensor]}
\end{equation}
with the appropriate contracted indices.
\begin{mmaIn}
DefTensor[alpha[-a],M,VertDeg$\to$1,PrintAs$\to$''$\alpha$'']\mmaSup\newline
DefMetric[-1,g[-a,-b],LCDer,{";","D"}]\mmaSup\newline
DefTensor[phi[],M,PrintAs$\to$''$\phi$'']\mmaSup\newline
DefTensor[xi[a],M,PrintAs$\to$''$\xi$'']\mmaSup\newline
VVFFromLieD[xi][{phi,g,alpha}]
\end{mmaIn}
\begin{mmaOut}
   $\mathcal{L}_\xi \alpha_{a} \wwedge \left(\frac{\delta}{\delta \alpha}\right)^{a} + \left(\frac{\delta}{\delta g}\right)^{ab} \mathcal{L}_\xi g_{ab} + \left(\frac{\delta}{\delta \phi}\right) \mathcal{L}_\xi \phi$
\end{mmaOut}

%/=-=-=-=-=-=-=\%
\subsection{Vertical operators}\label{subsection: Vertical operators}
%\=-=-=-=-=-=-=/%
\xCPS implements four vertical operators: \x{VertDiff}, \x{VertInt}, \x{VertLie}, and \x{VertBracket}. 
%/=*=\%
\subsubsection{\x{VertDiff}}
%\=*=/%
\x{VertDiff} is an operator of vertical degree $+1$ that takes the vertical exterior derivative of an expression. It automatically applies the following rules:
\begin{enumerate}
    \item It is linear and distributes over products using the Leibniz rule
    \[\dd(\alpha\wwedge\beta)=(\dd\alpha)\wwedge\beta+(-1)^{\x{VertDeg[$\alpha$]}}\alpha\wwedge(\dd\beta) \]
    \item Nilpotency: $\dd^2=0$.
    \item It commutes with \x{PD} (since it is non-dynamical).
    \item \x{VertDiff[$\tensor{\delta}{_{a}^{b}}$]=0} (since the Kronecker delta is non-dynamical).
    \item For a declared tensor \x{T}, \x{VertDiff[T]} returns the tensor $(\dd T)$.
\end{enumerate}
Concrete expansions, such as the variation of the Riemann tensor in terms of $\dd g$, are handled by \x{ExpandVertDiff} and explained in \cref{section: ExpandVertDiff}.

%/=*=\%
\subsubsection{\x{VertInt}}
%\=*=/%
\x{VertInt} is an operator of vertical degree $-1$ that takes the vertical interior product of an expression with a VVF. It automatically applies the following rules:
\begin{enumerate}
    \item It is linear and distributes over products using the Leibniz rule
    \[\ii_{\x{VVF}}(\alpha\wwedge\beta)=(\ii_{\x{VVF}}\alpha)\wwedge\beta+(-1)^{\x{VertDeg[$\ii_{\x{VVF}}$]VertDeg[$\alpha$]}}\alpha\wwedge(\ii_{\x{VVF}}\beta) \]
    \item $\ii_{\x{VVF}}\alpha=0$ if \x{VertDeg[$\alpha$]=0}.
\end{enumerate}
If \x{VVF} is an unresolved symbol, the output contains unevaluated terms of the form \x{VertInt[VVF]\texttt{@}$\tensor{T}{^a_b}$} When \x{VVF} is an explicit expression such as \cref{eq: VVF example}, the terms can be further simplified using \x{ExpandVertInt}, as we will see in \cref{section: ExpandVertInt}.

%/=*=\%
\subsubsection{\x{VertLie}}
%\=*=/%
\x{VertLie} is an operator of vertical degree $0$ that takes the vertical Lie derivative of an expression with respect to a VVF. It automatically applies the following rules:
\begin{enumerate}
    \item It is linear and distributes over products using the Leibniz rule
    \[\LL_{\x{VVF}}(\alpha\wwedge\beta)=(\LL_{\x{VVF}}\alpha)\wwedge\beta+(-1)^{\x{VertDeg[VVF]VertDeg[$\alpha$]}}\alpha\wwedge(\LL_{\x{VVF}}\beta) \]
    \item \x{VertLie[VVF][$\tensor{\delta}{_{a}^{b}}$]=0} (since the Kronecker delta is non-dynamical).
\end{enumerate}
Cartan's magic formula relates \x{VertLie} to \x{VertDiff} and \x{VertInt}, and is implemented via \x{SortVertOperators} as explained below.

%/=*=\%
\subsubsection{\x{VertBracket}}
%\=*=/%
\x{VertBracket} represents the vertical Lie bracket of \x{VVF1} and \x{VVF2}. It automatically applies the following rules:
\begin{enumerate}
    \item It is bilinear and satisfies the super-anticommutative property (lexicographical order):
    \[\llbracket \x{VVF2},\x{VVF1}\rrbracket=-(-1)^{\x{VertDeg[VVF1]VertDeg[VVF2]}}\llbracket\x{VVF1},\x{VVF2}\rrbracket\]
    \item \x{$\llbracket$VVF,VVF$\rrbracket$=0} if \x{VertDeg[VVF]} is even.
\end{enumerate}

%/=*=\%
\subsubsection{\x{SortVertOperators}, \x{VertCartanMagicFormula}, and \x{VertBracketToVertLie}}\label{subsubsection: SortVertOperators}
%\=*=/%
These three commands rewrite expressions containing nested vertical operators in terms of the others. To understand how, first notice that there are six possible super-commutators between $\LL$, $\ii$, and $\dd$, defined as:
\begin{equation}\label{eq: supercommutator}
    [P, Q] = PQ - (-1)^{\x{VertDeg[$P$]}\x{VertDeg[$Q$]}} QP
\end{equation}
Under appropriate conditions on the vector fields (they are prolongations of evolutionary vector fields, see for instance \cite[1.16,1.17,3.17,3.26]{anderson1989variational}), their result can be written in terms of $\{\LL, \ii, \dd\}$:
\[\begin{array}{lll}
    [\LL_{\mathtt{vvf1}}, \LL_{\mathtt{vvf2}}] = \LL_{\llbracket\mathtt{vvf1}, \mathtt{vvf2}\rrbracket} \qquad& 
    [\LL_{\mathtt{vvf1}}, \ii_{\mathtt{vvf2}}] = \ii_{\llbracket\mathtt{vvf1}, \mathtt{vvf2}\rrbracket} \qquad& 
    [\LL_{\mathtt{vvf}}, \dd] = 0  \\
    & 
    [\ii_{\mathtt{vvf1}}, \ii_{\mathtt{vvf2}}] = 0 &
    [\ii_{\mathtt{vvf}}, \dd] = \LL_{\mathtt{vvf}} \\
    & &
    [\dd, \dd] = 0 
\end{array}\]
The last one is $\dd^2=0$, which is implemented automatically. Whenever other operators are nested, we can sort them out using \x{SortVertOperators} according to a specified sorting order $\{P,Q,R\}$ (which can be parsed as an argument or defined globally via \x{\$SortVertOperatorsOrder}). The command \x{SortVertOperators} (analogous to \x{SortDerivationsRule} of \xTerior) applies \cref{eq: supercommutator} to choose the order $PQ$ or $QP$, and substitutes the supercommutator using the table above. The diagonal entries of the table are resolved by sorting \x{VVF1} and \x{VVF2} lexicographically.

Conversely, the LHS can be recovered from the RHS by applying \x{VertCartanMagicFormula} and \x{VertBracketToVertLie}. The former substitutes $\LL$ by applying Cartan's magic formula. Meanwhile, the latter rewrites \x{VertBracket} in terms of \x{VertLie} as follows (omiting indices for simplicity):
\begin{equation}
\llbracket T \wwedge  \left(\frac{\delta}{\delta R}\right) , S \wwedge \left( \frac{\delta}{\delta Q} \right)\rrbracket = \left( \LL_{T \left( \frac{\delta}{\delta R} \right)} S \right) \wwedge \left( \frac{\delta}{\delta Q} \right) - (-1)^{\|\x{vvf1}\|\|\x{vvf2}\|}\left(\LL_{S \left( \frac{\delta}{\delta Q} \right)} T \right) \wwedge \left( \frac{\delta}{\delta R} \right)
\end{equation}
where $\|\x{vvf1}\|$ is the \x{VertDeg} of the first VVF and analogously for $\|\x{vvf2}\|$. This is the infinite-dimensional analogue of the finite-dimensional formula:
\begin{equation}
[V^i \partial_i, W^j \partial_j]^k \partial_k = (V^m \partial_m W^k - W^m \partial_m V^k)\partial_k = (\mathcal{L}_V W^k)\partial_k - (\mathcal{L}_W V^k)\partial_k
\end{equation}

%/=-=-=-=-=-=-=\%
\subsection{Variational relations}
%\=-=-=-=-=-=-=/%
A fundamental aspect of \xCPS is its careful tracking of variational dependencies among tensors. For instance, the first variation of the Levi-Civita connection is given by:
\begin{equation}\label{eq: dlGamma=}
 \tensor{(\dd\Gamma)}{^a_b_c}=\frac{g^{ad}}{2}\Big(D_b(\dd g)_{dc}+D_c(\dd g)_{bd}-D_d(\dd g)_{cb}\Big)
\end{equation}
This is analogous to the chain rule in the finite-dimensional case: $\d f=\frac{\partial f}{\partial x^i}\d x^i$. We can see that the tensor $\Gamma$ depends variationally on $g$, in the sense that if $\dd g=0$, then $\dd\Gamma=0$. Thus, we can create a relation between the two. The converse does not hold: $\dd\Gamma=0$ is possible with $\dd g\neq0$, so the relation is directed $g\to\Gamma$.

Dependencies can be more intricate: consider a vector $v^a$ and its derived tensor $\tensor{T}{^a_b}:=D_bv^a$. It is well-known that
\begin{equation}\label{eq: dlCovD=}
    \tensor{(\dd T)}{^a_b}=D_b(\dd v)^a+\tensor{(\dd\Gamma)}{^a_b_c}v^c
\end{equation}
Hence, $T$ depends variationally on both $v$ and $\Gamma$, in the sense that if $\dd v=0$ and $\dd\Gamma=0$, then $\dd T=0$. We can then create the relations $v\to T\leftarrow \Gamma$. These examples show that tracking variational dependencies requires a directed graph. Furthermore, cyclic relations must be allowed. If two tensors are related by $T_{ab}=2S_{ab}$, then $S_{ab}=\frac{1}{2}T_{ab}$, establishing both $T\to S$ and $S\to T$. While \xAct has a basic built-in relation checker (\x{ImplicitTensorDepQ}), \xCPS introduces a dedicated global object, \x{\$VariationalGraph}. This graph encodes all defined tensors as vertices and their variational dependencies as directed edges

\begin{remark}
Variational relations involving tensors with a non-zero \x{VertDeg} exist but cannot be currently encoded in \xCPS. For example, given $T,R,S$ with \x{VertDeg=0} and the $1$-form given by 
\[\omega=T(\dd R+\dd S)\qquad\longrightarrow\qquad\dd\omega=\dd T\wwedge (\dd R+\dd S)\]
We see that $T\to\omega$, but we do not have $R\to\omega$ or $S\to\omega$. However, an indirect relationship exists (if $R=-S$, then $\dd\omega=0$), but it cannot be modeled via simple graph edges. Future versions of \xCPS might include these relations through higher order chains.
\end{remark}

Many variational relations are implemented in \xCPS. For instance, when \x{DefMetric} or \x{DefCovD} are called, the obvious relations are created (\x{Riemann} depends on the metric, \x{Ricci} depends on \x{Riemann} and the metric, and so on). This data can be queried using \x{VariationalRelationsOf[tensor]}, which returns the labeled subgraph of \x{\$VariationalGraph} connected to \x{tensor}. The argument can be a list of tensors or \x{All} to display the complete graph.
\begin{mmaIn}
VariationalRelationsOf[RiemannLCDer]
\end{mmaIn}
\begin{mmaOut}
\mmaGraphicOut[width=.5\linewidth]{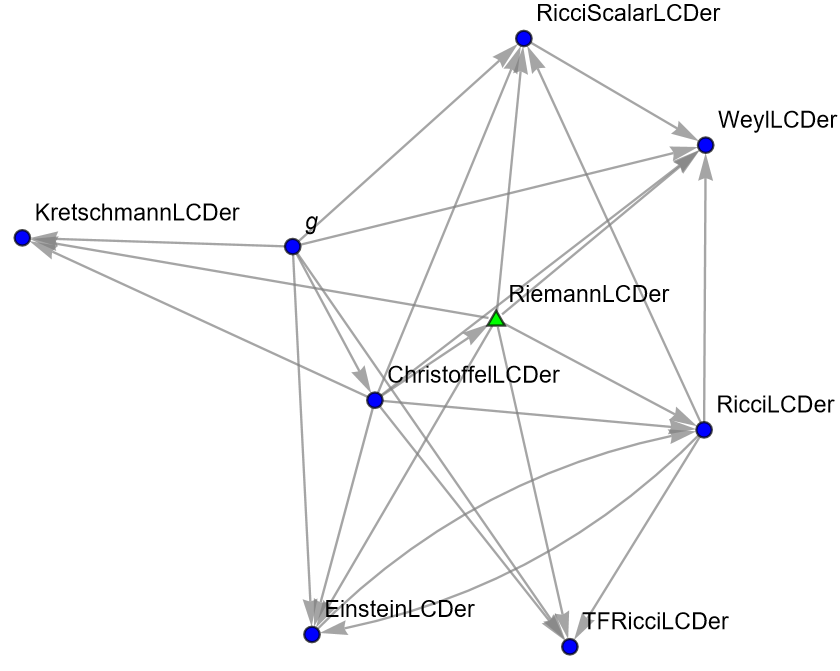
}
\end{mmaOut}

Users can add or remove relations using \x{AddVariationalRelation[tensor1$\to$tensor2]} and \x{RemoveVariationalRelation[tensor1$\to$tensor2]}.

\begin{mmaIn}
DefTensor[\{v[a],w[a],u[a],A[-a],B[]\},M]\newline
AddVariationalRelation[v$\to$w$\to$u]\newline
AddVariationalRelation[A$\to$B$\to$w]
\end{mmaIn}

\x{AddVariationalRelation} adds only the abstract relation without specifying the formula; this is done via \x{GenerateExpandVertDiffRule} (\cref{section: ExpandVertInt}). Additionally, \x{VariationalRelationsOf} accepts the option \x{Directed$\to$Out} or \x{Directed$\to$In} that restricts, respectively, to outbound or inbound dependencies. 

\begin{mmaIn}
VariationalRelationsOf[w]\newline
VariationalRelationsOf[w,Directed$\to$In]
\end{mmaIn}
\begin{mmaOut}
\mmaGraphicOut[width=.48\linewidth]{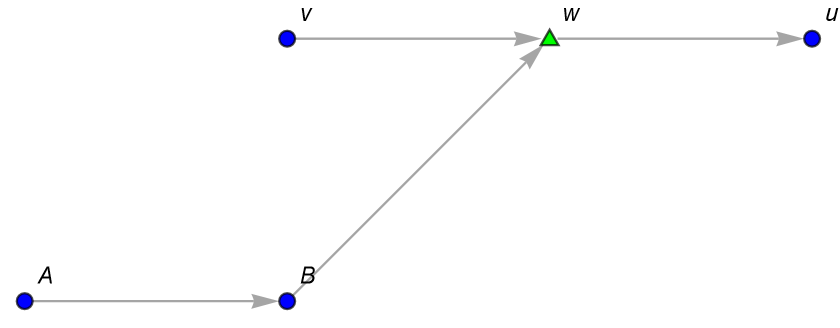}\\
\mmaGraphicOut[width=.42\linewidth]{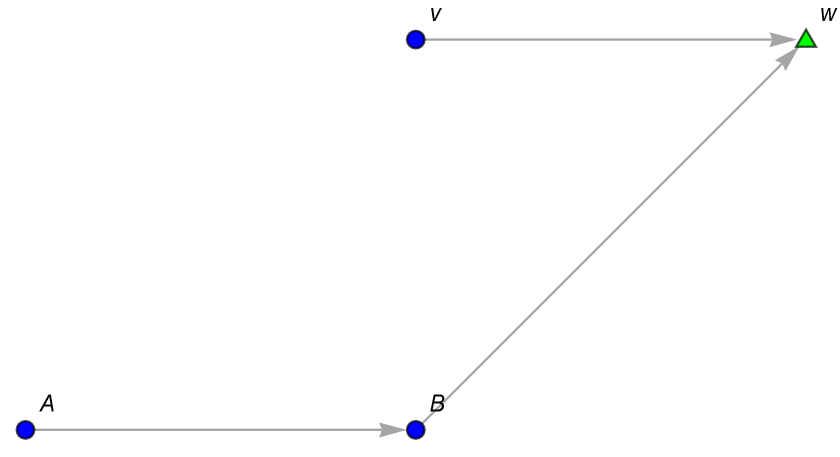}
\end{mmaOut}
 
The \x{ConstantTensors} option can be passed to \x{VariationalRelationsOf} to declare specific tensors as variationally constant. Once initial constant tensors are seeded, the ``constantness'' is propagated forward through the graph with backward pruning: a node is constant if all the incoming edges are constant. This recursive propagation algorithm is a major technical novelty in \xCPS.

Constant nodes are rendered as red squares and relations between them as red edges. The central node of the graph (the tensor passed as argument) is always drawn as a green triangle for visibility, with its dynamical status determined by its surrounding edges.\vspace{1ex}

\begin{mmaIn}
VariationalRelationsOf[w,ConstantTensors$\to$\{B\}]\newline
VariationalRelationsOf[w,ConstantTensors$\to$\{B,v\}]
\end{mmaIn}
\begin{mmaOut}
\mmaGraphicOut[width=.5\linewidth]{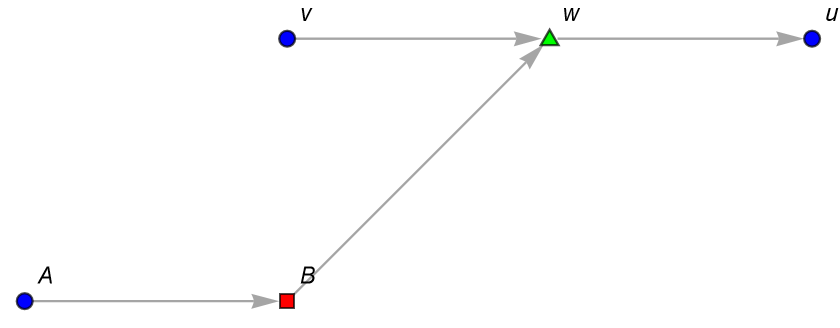}\\
\mmaGraphicOut[width=.5\linewidth]{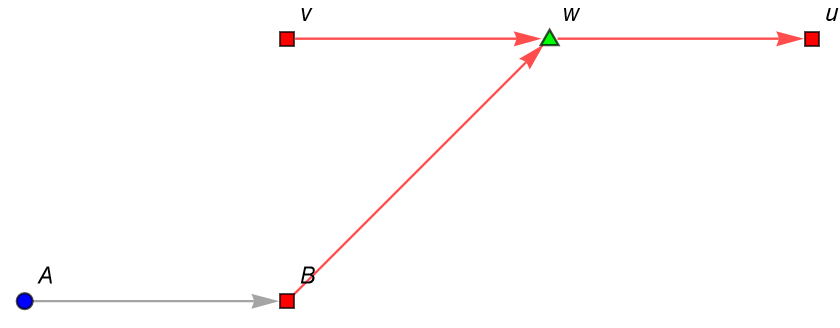}
\end{mmaOut}
This dependency graph is essential for computing variational derivatives, since only the dynamical fields must be taken into account. Several commands presented on \cref{sec: variational calculus} make use of the function \x{VariationallyConstantQ[ConstantTensors,\{\}][tensor]}, which returns \x{True} if \x{tensor} is variationally constant (red square in the graph) and \x{False} otherwise.

%/=-=-=-=-=-=-=\%
\subsection{ExpandVertDiff}\label{section: ExpandVertDiff}
%\=-=-=-=-=-=-=/%
\x{ExpandVertDiff[][expr]} expands the vertical exterior derivative of \x{expr}. It operates analogously to \x{ExpandPerturbation} from \xPert \cite{xPert}, but incorporates significant additional capabilities. 

We can use \x{GenerateExpandVertDiffRule[$\{$dltensor,expansion$\}$,options]} to generate the expansion rule \x{dltensor$\to$expansion} to be used by \x{ExpandVertDiff}. A rule can be removed with \x{RemoveExpandVertDiffRule[dltensor]}.

\begin{mmaIn}
    DefTensor[beta[-a,-b],M,VertDeg$\to$2,PrintAs$\to$"$\beta$"]\mmaSup\newline
    GenerateExpandVertDiffRule[{dlbeta[-a,-b],dlalpha[-a]$\wwedge\,$alpha[-b]}]\mmaSup\newline
    dlbeta[-a,-b] == (dlbeta[-a,-b] // ExpandVertDiff[])
\end{mmaIn}
\begin{mmaOut}
$(\dd \beta)_{ab} = (\dd \alpha)_{a} \wwedge \alpha_{b}$
\end{mmaOut}
\x{ExpandVertDiff[]} is applied recursively. Note that \xCPS does not perform infinite-loop checks.
\begin{mmaIn}
GenerateExpandVertDiffRule[{dlalpha[-a],alpha[b]$\wwedge$\,dlg[-a,-b]}]\mmaSup\newline
dlbeta[a,-b] == (dlbeta[a,-b] // ExpandVertDiff[])
\end{mmaIn}
\begin{mmaOut}
$(\dd \beta)_{ab} = g^{cd} \alpha_{d} \wwedge (\dd g)_{ac} \wwedge \alpha_{b}$
\end{mmaOut}
\x{ExpandVertDiff} expands expressions involving \x{LieD}, \x{CovD}, and \x{Bracket} using predefined rules, such as \cref{eq: dlCovD=}. Most of these rules can be toggled off via options like \x{ExpandVertDiffLieD}, \x{ExpandVertDiffCovD}, and \x{ExpandVertDiffBracket}, which are all \x{True} by default.
\begin{mmaIn}
VertDiff[LCDer[-a][alpha[-b]]] == (VertDiff[LCDer[-a][alpha[-b]]] \\//  ExpandVertDiff[] //ContractMetric // Simplification)
\end{mmaIn}
\begin{mmaOut}
$\dd [D_{a}\alpha_{b}] = \tfrac{1}{2} (3 \alpha^{c} \wwedge D_{a}(\dd g)_{bc} + \alpha^{c} \wwedge D_{b}(\dd g)_{ac} -  \alpha^{c} \wwedge D_{c}(\dd g)_{ab} + 2 D_{a}\alpha^{c} \wwedge (\dd g)_{bc})$
\end{mmaOut}
Finally, \x{ExpandVertDiff} also handles tensors defined alongside a CovD or metric, such as \x{Riemann}, \x{Ricci}, and \x{Torsion}, using predefined rules like \cref{eq: dlGamma=}:
\begin{mmaIn}
VertDiff[ChristoffelLCDer[a,-b,-c]] == (VertDiff[ChristoffelLCDer[a,-b,-c]]  \\ // 
     ExpandVertDiff[] )
\end{mmaIn}
\begin{mmaOut}
$(\dd \Gamma [D])^{a}{}_{bc} = \tfrac{1}{2} g^{ad} (D_{b}(\dd g)_{dc} + D_{c}(\dd g)_{bd} -  D_{d}(\dd g)_{bc})$
\end{mmaOut}
One of the most interesting features of \x{ExpandVertDiff} is the option \x{HoldExpandVertDiff}, which acts as a circuit breaker to prevent deeper recursive expansion. For instance, it allows one to expand the variation of the Riemann tensor in terms of $\dd\Gamma$ rather than unwinding it down to $\dd g$:
\begin{mmaIn}
expr = VertDiff[RiemannLCDer[-a,-b,-c,-d]];\mmaSup\newline
expr // ExpandVertDiff[] // ContractMetric // Simplification\newline
expr // ExpandVertDiff[HoldExpandVertDiff$\to$ChristoffelLCDer] \\// ContractMetric // Simplification
\end{mmaIn}
\begin{mmaOut}
$\tfrac{1}{2} (- D_{a}D_{b}(\dd g)_{c}{}^{d} -  D_{a}D_{c}(\dd g)_{b}{}^{d} + D_{a}D^{d}(\dd g)_{bc} + D_{b}D_{a}(\dd g)_{c}{}^{d} + D_{b}D_{c}(\dd g)_{a}{}^{d} -  D_{b}D^{d}(\dd g)_{ac})$\\
$D_{b}(\dd \Gamma [D])^{d}{}_{ac}- D_{a}(\dd \Gamma [D])^{d}{}_{bc} $
\end{mmaOut}
Another essential option is \x{ConstantTensors}, which indicates the list of tensors to be considered as constant (non-dynamic). As explained in the previous section for \x{VariationalRelationsOf}, these constant tensors might force other tensors to be constant.
\begin{mmaIn}
expr = VertDiff[RicciScalarLCDer[]];\mmaSup\newline
expr // ExpandVertDiff[] // ContractMetric // Simplification\newline
expr // ExpandVertDiff[HoldExpandVertDiff$\to$RiemannLCDer]\\ // 
  ContractMetric // Simplification\newline
expr // ExpandVertDiff[ConstantTensors$\to$RiemannLCDer] // ContractMetric \\// Simplification
\end{mmaIn}
\begin{mmaOut}
$- (\dd g)^{ab} R[D]_{ab} + D_{b}D_{a}(\dd g)^{ab} -  D_{b}D^{b}(\dd g)^{a}{}_{a}$\\
$(\dd R[D])^{ab}{}_{ab} - 2 (\dd g)^{ab} R[D]_{ab}$\\
$-2 (\dd g)^{ab} R[D]_{ab}$
\end{mmaOut}
The option \x{NonConstantTensors} is the opposite of \x{ConstantTensors}, providing the list of dynamical tensors by enforcing every other tensor to be constant.

%/=-=-=-=-=-=-=\%
\subsection{ExpandVertInt}\label{section: ExpandVertInt}
%\=-=-=-=-=-=-=/%
\x{ExpandVertInt} behaves to \x{VertInt} as \x{ExpandVertDiff} behaves to \x{VertDiff}, with the additional rule explained when we introduced \cref{eq: XX=int delta delta}:
\begin{equation}\label{eq: VertInt VVF_ij}
\ii_{\x{coeff[inds1]} \wwedge \frac{\delta}{\delta \x{tensor}_i}[\x{-inds1}]]}\Big[\dd \x{tensor}_j[\x{inds2}]\Big]=
\begin{array}{|ll}
0            & \text{if } i\neq j\\
\x{coeff[inds2]} & \text{if } i=j
\end{array}
\end{equation}
This is the infinite-dimensional version of
\begin{equation}
    \iota_{f(\vec{x})\frac{\partial}{\partial x^i}}\Big[\d x^j\Big]=
\begin{array}{|ll}
0            & \text{if } i\neq j\\
f(\vec{x}) & \text{if } i=j
\end{array}
\end{equation}
Before applying the rule of \cref{eq: VertInt VVF_ij}, \x{ExpandVertInt} calls \x{ExpandVertDiff} with the option \x{HoldExpandVertDiff} active for the field \x{tensor}${}_i$, the unique tensor whose variational vector $\frac{\delta}{\delta \x{tensor}_i}$ appears in the 
VVF. This ensures that the matching condition of \cref{eq: VertInt VVF_ij} is checked before 
$\dd\x{tensor}_j$ is expanded into more primitive variations.

\begin{remark}
    The VVF knows how to act over $\dd$\x{tensor} (vertically exact forms). If we want to define a VVF acting over non-exact vertical $k$-forms, we need \x{DefGeneralizedVVF} but, for brevity, we do not discuss it further here and refer the interested reader to the \xCPS documentation.
\end{remark}

%/=-=-=-=-=-=-=\%
\subsection{Divergences and reconstruction of potentials}
%\=-=-=-=-=-=-=/%
In this subsection, we introduce two of the most powerful tools of \xCPS, both of 
independent interest beyond the CPS formalism: \x{DivergenceQ} and \x{FindPotentialGradient}. The former checks if an expression is a total divergence $\x{expr=}\nabla_a V^a$ and the latter attempts to find the potential.

 \newpage

\begin{remarks}
    \item \x{FindPotentialGradient} is a heuristic search that can be applied to any tensorial expressions (with or without free indices).
    \item \x{DivergenceQ} is a true algorithm, but is currently restricted to scalar expressions (no free indices). A heuristic extension to support free indices is under development.
\end{remarks}

%/=*=\%
\subsubsection{DivergenceQ}\label{subsubsection: divergenceQ}
%\=*=/%
\x{DivergenceQ} checks whether a scalar expression gives rise to a null Lagrangian, i.e., a Lagrangian with no EOMs. Under certain assumptions (see \cref{subsection: Ambiguities} and \cite{margalef2021geometric}), this is equivalent to being a divergence. \x{DivergenceQ} requires an auxiliary metric to form a Lagrangian density but it can check if an expression is a divergence with respect to non-Levi-Civita CovDs.
\begin{itemize}
    \item \x{DivergenceQ[metric][expr]} checks if \x{expr=$\nabla_aV^a$} for some vector $V^a$ where $\nabla$ is the Levi-Civita connection of \x{metric}.
    \item \x{DivergenceQ[metric,der][expr]} checks if \x{expr=$D_aV^a$} for some vector $V^a$ where \x{D=der} is a CovD (not necessarily the Levi-Civita of metric). An auxiliary metric is required.
    \item \x{DivergenceQ[\{metric,LC\},der][expr]} checks if \x{expr=$D_a V^a$} for some $V^a$ defined over an inner bundle (it might have contracted inner indices) where \x{D=der} is a CovD (not necessarily the Levi-Civita of \x{metric}). An auxiliary \x{metric} and an extended CovD \x{LC} are required.
\end{itemize}
The second and third cases are similar to the first one with some additional technicalities that are explained in the documentation. Thus, for conciseness, it is enough to understand the first case. \x{DivergenceQ} proceeds by forming a Lagrangian density by multiplying by the appropriate power of \x{Detmetric[]}, and checking that all resulting EOMs vanish. If they are trivially zero, \x{DivergenceQ} returns \x{True}.
\begin{mmaIn}
    expr1 = LCDer[-a][xi[a] phi[]]\newline
    DivergenceQ[g][expr1]
\end{mmaIn}
\begin{mmaOut}
    $\xi^{a} D_{a}\phi + \phi D_{a}\xi^{a}$\\
    True
\end{mmaOut}
Sometimes, complicated expressions are not trivially zero without further simplifications.
\begin{mmaIn}
    expr2 = LCDer[-a]@LCDer[-b][LCDer[a][xi[b]phi[]]-LCDer[b][xi[a]phi[]]]\newline
    DivergenceQ[g][expr2]\newline
    DivergenceQ[g][expr2] // SortCovDs // Simplification
\end{mmaIn}
\begin{mmaOut}
    $\xi^{b} D_{a}D_{b}D^{a}\phi + \phi D_{a}D_{b}D^{a}\xi^{b} + D_{a}D_{b}\xi^{b} D^{a}\phi - D^{a}\phi D_{b}D_{a}\xi^{b} -  \xi^{b} D_{b}D_{a}D^{a}\phi -  \phi D_{b}D_{a}D^{a}\xi^{b}$\\
    $-2 D_{a}D_{b}D^{b}\xi^{a} + 2 D_{b}D_{a}D^{b}\xi^{a} = 0$\\
    True
\end{mmaOut}
If the user is confident that \x{expr} is not a divergence and requires a boolean \x{False}, they can enable \x{CheckZero} (\x{False} by default), which enforces strict equality, \x{===}, instead of \x{==}. Use this option with caution, as it may return \x{False} even when further simplifications would yield \x{True}.

\begin{mmaIn}
    NonDivergence = phi[] LCDer[-a][xi[a]]\newline
    DivergenceQ[g][NonDivergence]\newline
    DivergenceQ[g,CheckZero$\to$True][NonDivergence]
\end{mmaIn}
\begin{mmaOut}
    $\phi D_{a}\xi^{a}$\\
    $- \tfrac{1}{2} g^{ab} \xi^{c} D_{c}\phi = 0$ \&\& $- D_{a}\phi = 0$ \&\& $D_{a}\xi^{a} = 0$\\
    False
\end{mmaOut}

%/=*=\%
\subsubsection{FindPotentialGradient}\label{subsubsection: FindPotentialGradient}
%\=*=/%
 \x{FindPotentialGradient[der][expr]} attempts to find a tensorial expression \x{potential} such that \x{expr=der[ind][potential]}. This is the ``inverse'' of the Leibniz rule. This is a genuinely difficult problem in general. Even in the well-studied case of differential forms, knowing whether a $k$-form $\alpha$ is exact, $\d\beta=\alpha$, leads to the theory of cohomology. However, in that case there is an additional condition that the potential $\beta$ should be a $(k-1)$-form (a totally antisymmetric tensor). That forces $\alpha$ to be closed, i.e. $\d\alpha=0$ since $\d^2=0$. For general tensors we do not have that constraint but there is no known algorithm to obtain such potential. The procedure is heuristic: it either terminates with a potential or fails to converge (because it gets into a loop, which is automatically detected by the algorithm, or because it does not exist). The procedure can be summarized as follows:
 \begin{enumerate}
     \item Rewrite all \x{Brackets}, \x{LieD} and CovDs in terms of \x{der}.
     \item Turn \x{expr} into a list \x{\{monomial${}_1$,\ldots,monomial${}_N$\}} where each element is a monomial.
     \item For each monomial, split it into two factors: the term with the most nested \x{der} applications (\x{highestder}) and the remainder (\x{rest}), recording the two depth values:
     \[\x{\{\{H${}_1$,h${}_1$\},\{highestder${}_1$,rest${}_1$\}\},\ldots,\{\{H${}_N$,h${}_N$\},\{highestder${}_N$,rest${}_N$\}\}}\]
     \x{H}${}_i$ is the number of nested \x{der} in \x{highestder}${}_i$ and \x{h}${}_i$ the second highest depth value, used to order monomials when \x{H}${}_i$ = \x{H}${}_j$.
     \item Order the monomials using \x{\{H${}_i$,h${}_i$\}}. Let us assume they are already ordered.
     \item Apply the inverse of the Leibniz rule to the monomial with the highest number of \x{der}. If \x{highestder${}_1$=der[-a][term${}_1$]}:
     \[\x{\{highestder${}_1$,rest${}_1$\}}\quad\longrightarrow\quad\Big(\x{der[-a][term${}_1$rest${}_1$]},\x{\{term${}_1$,-der[-a][rest${}_1$]\}}\Big)\]
     The first term forms part of \x{potential} (with a wrapper to prevent \xAct from re-expanding it via the Leibniz rule). The second element is a new pair $\{\x{term}_1, -\x{der[-a][rest}_1\x{]}\}$, representing the remaining expression to be processed: the first element is the term inside the derivative, while the second element is the remainder of extracting the first via Leibniz rule.
     \item[5'.] There is an alternative step 5 for certain cases, as explained below.
     \item WWedge-multiply the pairs and add all the monomials:
     \[-\x{term${}_1\wwedge\ $der[-a][rest${}_1$]+highestder${}_2\wwedge\ $rest${}_2$ + $\cdots$ + highestder${}_N\wwedge\ $rest${}_N$}\]
     \x{der[-a][rest${}_1$]} is expanded via the Leibniz rule and some cancellations may occur.
     \item Apply the whole procedure again, storing in each step a new term for \x{potential}.
     \item The procedure finishes if the expression of the 6th step is zero. Then, the accumulated \x{potential} is returned.
 \end{enumerate}
\begin{mmaIn}
expr = LCDer[-a][LCDer[-b]@v[b]LCDer[-c]@alpha[-d]+alpha[-d]$\wwedge$ alpha[-c]]\newline
FindPotentialGradient[LCDer][expr]
\end{mmaIn}
\begin{mmaOut}
    $\alpha_{c} \wwedge D_{a}\alpha_{d} + D_{a}\alpha_{c} \wwedge \alpha_{d} + D_{a}D_{c}\alpha_{d} D_{b}v^{b} + D_{a}D_{b}v^{b} D_{c}\alpha_{d}$\\
    $(n_{{}_D})_{a} (\alpha_{c} \wwedge \alpha_{d} + D_{b}v^{b} D_{c}\alpha_{d})$
\end{mmaOut}

\begin{remark}\label{remark: n_CovD}
The previous output is expressed using the multipurpose symbol $(n_{{}_D})_{a}$, which has a dual interpretation. First, within the spacetime manifold, it can be interpreted as a placeholder to contract the free index of the previous expressions to turn the vectors (see \cref{subsection: top-forms vs densities}) into scalars, which are easier to handle. However, the most important interpretation is as the unitary normal and future directed vector field to a Cauchy hypersurface $\Sigma$ (see \cref{eq: Omega=n_a U^a vol Sigma}).
\end{remark}
 
Using the term with most \x{der} on the 5th step ensures that the numbers \{\x{H}${}_i$,\x{h}${}_i$\} are reduced if \x{H}${}_i$>\x{h}${}_i+1$, otherwise, it might lead to a loop. This situation arises for expressions like $\psi\nabla_a\phi$ \x{H}${}_i$=\x{h}${}_i+1$) or $\nabla_a\psi\nabla_b\phi$ (\x{H}${}_i$=\x{h}${}_i$), which are not gradients. Importantly, even genuine gradients can lead to a loop. For instance, consider $\phi\nabla_a \phi$:
\[\phi\nabla_a \phi\quad\longmapsto\quad\{\{1,0\},\{\nabla_a\phi,\phi\}\}\quad\longmapsto\quad\Big(\nabla_a(\phi^2),\{\phi,-\nabla_a\phi\}\Big)
\]
The \x{potential} so far is $\phi^2$ and the new expression is $-\phi\nabla_a\phi$, which is minus the original expression. Applying the procedure again leads to a second sign flip, so we recover the original expression with a vanishing accumulated potential. This is a 2-step loop even though the initial expression is equal to $\nabla_a(\frac{1}{2}\phi^2)$. Likewise, a similar thing would happen if we start with $v^a\nabla_bv_a=\nabla_b(\frac{1}{2}v^av_a)$. To handle these cases, we apply step 5':

\begin{enumerate}
    \item[5'.] Check if the tensor appearing inside the CovD appears with the opposite indices on \x{rest}. If so, perform the ``square-Leibniz'' rule, otherwise, perform the usual Leibniz rule (step 5).
\end{enumerate}
\begin{mmaIn}
expr = phi[]LCDer[-a][phi[]]\newline
FindPotentialGradient[LCDer][expr]
\end{mmaIn}
\begin{mmaOut}
    $\phi D_{a}\phi$\\
    $\tfrac{1}{2}(n_{{}_D})_{a} \phi^2$
\end{mmaOut}
\begin{mmaIn}
expr = v[b]LCDer[-a][v[-b]]\newline
FindPotentialGradient[LCDer][expr]
\end{mmaIn}
\begin{mmaOut}
    $v^{b} D_{a}v_{b} + v_{b} D_{a}v^{b}$\\
    $\tfrac{1}{2}(n_{{}_D})_{a} v_{b} v^{b}$
\end{mmaOut}

 \begin{remark}
      \x{FindPotentialGradient} accepts two optional arguments: a positive integer \x{iteration}, that limits the amount of iterations to obtain the middle steps, and \x{optionalfunctions}, that are applied between steps 6 and 7 to rewrite the expression and help with the convergence of the procedure. We will see in \cref{subsection: sf and GR} an example of their need.
 \end{remark}

%%%%%%%%%%%%%%%%%%%%%%%%%%%%%%%%%%%%%%%%%%%%%%%%%%%%%%%%%%%%%%%%%%%%%%%%%%%%%%%%%%%%%%%%%%
%%%%%%%%%%%%%%%%%%%%%%%%%%%%%%%%%%%%%%%%%%%%%%%%%%%%%%%%%%%%%%%%%%%%%%%%%%%%%%%%%%%%%%%%%%
%%%%%%%%%%%%%%%%%%%%%%%%%%%%%%%%%%%%%%%%%%%%%%%%%%%%%%%%%%%%%%%%%%%%%%%%%%%%%%%%%%%%%%%%%%

%/=====================\%
\section{Variational calculus}\label{sec: variational calculus}
%\=====================/%
This section presents the variational calculus functions of \xCPS, applied to a Lagrangian density.

%/=-=-=-=-=-=-=\%
\subsection{First variation}
%\=-=-=-=-=-=-=/%
The command \x{FirstVariation[tensors,der][L]} computes the first variation in the form of \cref{eq: first variation densities} with \x{tensors} as the dynamical fields (if empty, all are assumed dynamic) and \x{der} is the CovD used to perform the Leibniz rule (it can be omitted if \x{L} contains only one CovD). The procedure leading to \cref{eq: first variation densities} is:

\begin{enumerate}
    \item Take \x{VertDiff[L]} followed by \x{ExpandVertDiff[NonConstantTensors$\to$tensors]} which, schematically, takes the form
    \begin{equation}
       \sum_{i=1}^N\sum_{j=0}^{m_i}(T_{ij})^{a_1\cdots a_j}D_{a_1}\cdots D_{a_j}\dd\Phi_i       
    \end{equation}
    where $\{\Phi_i\}$ are the dynamical fields (possibly with upper and lower indices contracted with $T_{ij}$ that we omit for simplicity) and $j=0$ corresponds to no CovD acting on $\dd\Phi_i$.
    \item Apply the Leibniz rule to move all CovDs off $\dd\Phi_i$, yielding
    \begin{equation}      \sum_{i=1}^N\x{EOM}_i\dd\Phi_i+D_b\Big[\sum_{i=1}^N\sum_{j=0}^{m_i}\tensor{(R_{ij})}{^b^{a_1}^{\cdots}^{a_j}}D_{a_2}\cdots D_{a_j}\dd\Phi_i\Big] 
    \end{equation}
    \item To prevent the expansion of the last term, we wrap it into \x{TotalDerivativeOfder}, a new linear \x{InertHead}. To produce a scalar inside the \x{InertHead}, the free index \x{b} is contracted with the 1-form \x{NormalOfCovDder} (denoted $n_{{}_D}$, see \cref{remark: n_CovD}). This finally leads to
        \begin{equation}\label{eq: dlLagrangian split xCPS}   \sum_{i=1}^N\x{EOM}_i\dd\Phi_i+\x{TotalDerivativeOfCovDder}\Big[(n_{{}_D})_b\sum_{i=1}^N\sum_{j=0}^{m_i}\tensor{(R_{ij})}{^b^{a_1}^{\cdots}^{a_j}}D_{a_2}\cdots D_{a_j}\dd\Phi_i\Big] 
    \end{equation}
\end{enumerate}

The key step is the correct application of the Leibniz rule, which we now describe in more detail.

\begin{remark}
    \xAct implements the Leibniz rule (variational derivative) with \x{VarD} in a straightforward but not always efficient way. Roughly speaking, given a monomial, \x{VarD} acts on each factor by removing all derivatives in front of it via the Leibniz rule, discarding the total derivative:
\[(DD T^{ab})(D S_{ab})\quad\overset{\x{VarD}}{\longrightarrow}\quad-(D T^{ab})(DD S_{ab})\quad\overset{\x{VarD}}{\longrightarrow}\quad T^{ab}(DDD S_{ab})\]
The base case is reached when no derivatives remain: \x{VarD} checks whether the bare tensor is the one being varied, returning the rest of the expression if so, and zero otherwise. The same procedure is then applied to each remaining factor of the monomial:
\[(DD T^{ab})(D S_{ab})\quad\overset{\x{VarD}}{\longrightarrow}\quad-(DDD T^{ab})(S_{ab})\]
This approach has obvious disadvantages:
\begin{enumerate}
    \item \x{VarD} cannot know if it is removing the derivatives from the correct tensor until it finishes. If the wrong tensor is reached, the entire branch is discarded. This is particularly inefficient for higher order theories as we will see at the end of this subsection.
    \item There is no easy way to keep track of the total derivative terms that appear. For each step, we have one total derivative term but it might be discarded at the end.
    \item \x{VarD} does not take into account variational relations. For instance, \begin{mmaIn}
        \x{VarD[g[-a,-b],LCDer][RicciScalarLCDer[]]} 
    \end{mmaIn}
\begin{mmaOut}
$0$
\end{mmaOut}
    returns zero which is obviously not true. Traditionally, this computation was done using \x{xPert}, by computing the first order perturbation of the Lagrangian and then using \x{VarD} with respect to the first order perturbation of the tensor. This procedure is non-trivial for new users and additional variational relations cannot be incorporated. Moreover, the first two aforementioned problems still remain.
\end{enumerate}
\end{remark}
Since \x{VertDiff} and \x{ExpandVertDiff} guarantee that the expression is linear in $\dd\x{tensor}_i$ for each dynamical field \x{tensor}${}_i$, the Leibniz rule is never wasted and the total derivative can be safely kept. This is done with the private command \x{LeibnizRule}, which is similar to \x{VarD} but with a couple of changes:
\begin{enumerate}
    \item It accepts the option \x{KeepDivergence} (\x{True} by default) to keep the total derivative terms.
     \item It also applies the Leibniz rule to \x{LieD} expressions.
\end{enumerate}
\begin{mmaIn}
    LScl = $\frac{\x{1}}{\x{2}}$g[a,b] LCDer[-a]@phi[] LCDer[-b]@phi[] Sqrt[-Detg[]];\mmaSup\newline
FirstVariation[phi][LScl]\newline
FirstVariation[][LScl]
\end{mmaIn}
\begin{mmaOut}
$\x{TotalDerivativeOfLCDer}\bigl[\sqrt{-\dbtilde{g}}(\dd \phi) (n_{{}_D})^{a} D_{a}\phi \bigr] - \sqrt{-\dbtilde{g}}(\dd \phi) D_{b}D^{b}\phi$\\
$\x{TotalDerivativeOfLCDer}\bigl[\sqrt{-\dbtilde{g}}(\dd \phi) (n_{{}_D})^{a} D_{a}\phi \bigr] -  \sqrt{-\dbtilde{g}}(\dd \phi) D_{b}D^{b}\phi +{}$\newline$
(\dd g)_{ab} \bigl(- \tfrac{1}{2}\sqrt{-\dbtilde{g}}\,D^{a}\phi D^{b}\phi + 
\tfrac{1}{4} \sqrt{-\dbtilde{g}}(\dd \phi) g^{ab} D_{c}\phi D^{c}\phi \bigr)$
\end{mmaOut}

%/=-=-=-=-=-=-=\%
\subsection{EOM}
%\=-=-=-=-=-=-=/%
The command \x{EOM[tensor][L]} is similar to \x{FirstVariation} with the following differences:
\begin{enumerate}
    \item \x{tensor} is the only dynamical field (to obtain the EOM with respect to that field).
    \item  \x{KeepDivergence} is set to \x{False}, since the total derivative term is not important.
    \item The result usually carries free indices since the $\dd$\x{tensor} factor is dropped.
\end{enumerate}

The efficiency gain over \x{VarD} is most pronounced for higher-order theories:
\begin{mmaIn}
    Lhigh = LCDer[-a]@LCDer[-c]@RiemannLCDer[a,b,c,d] \\ LCDer[-b]@
   LCDer[-d]@phi[] Sqrt[-Detg[]]\newline   
    Timing[EOM[phi][Lhigh]]
\end{mmaIn}
\begin{mmaOut}
$\sqrt{-\dbtilde{g}}\,D_{a}D_{c}R[D]^{abcd} D_{d}D_{b}\phi $\\
    $\{0.078125,\sqrt{-\dbtilde{g}}\,g^{ae} g^{bf} g^{ch} g^{di} D_{b}D_{d}D_{a}D_{c}R[D]_{efhi}\}$
\end{mmaOut}
\begin{mmaIn}
<{}< xAct\`{}xPert\`{}\mmaSup\newline
DefMetricPerturbation[g,pertg,$\varepsilon$]\mmaSup\newline
DefTensorPerturbation[pertphi[LI[1]],phi[],M]\mmaSup\newline
Timing[VarD[pertphi[LI[1]],LCDer](Perturbation[Lhigh,1] \\ // ExpandPerturbation)]]
\end{mmaIn}
\begin{mmaOut}
    $\{0.1875,\delta_{1}{}^{1} \sqrt{-\dbtilde{g}}\,D_{d}D_{b}D_{c}D_{a}R[D]^{abcd}\}$
\end{mmaOut}
For simpler theories, the overhead of \x{VertDiff} and \x{ExpandVertDiff} may offset the savings; in those cases, however, both methods are essentially instantaneous.

\begin{remarks}
    \item In the previous examples, the EOMs agree (up to index contraction and index relabeling) but the computation time is twice as long with \xPert. \xPert also introduces a factor $\tensor{\delta}{_1^1}=1$, which is a trace of the perturbation parameter. 
    \item The difference is noticeable when there are multiple fields and the EOM is taken with respect to the field with fewer derivatives. There, \x{VarD} performs unnecessary Leibniz rule applications on the remaining fields.
    \item \xPert and \xCPS are very similar in this regard, but \xCPS is simpler for the user: \x{EOM} requires no additional setup beyond defining the Lagrangian.
\end{remarks}

%/=-=-=-=-=-=-=\%
\subsection{EnergyMomentum}\label{subsection: EnergyMomentum}
%\=-=-=-=-=-=-=/%
\x{EnergyMomentum} computes the energy-momentum tensor of a Lagrangian, i.e., up to a $-2/\sqrt{-\det g}$ factor, the EOM with respect to the main metric (the first defined metric in \xAct).
\begin{mmaIn}
EnergyMomentum[LScl] // ContractMetric // Simplification\newline
-$\frac{\x{2}}{\sqrt{\x{-Detg[]}}}$EOM[g][LScl] // ContractMetric // Simplification
\end{mmaIn}
\begin{mmaOut}
    $D^{a}\phi D^{b}\phi -  \tfrac{1}{2} g^{ab} D_{c}\phi D^{c}\phi$\\
$D^{a}\phi D^{b}\phi -  \tfrac{1}{2} g^{ab} D_{c}\phi D^{c}\phi $
\end{mmaOut}

%/=-=-=-=-=-=-=\%
\subsection{SymplecticPotential and SymplecticCurrent}\label{subsection: symplectic potential and current}
%\=-=-=-=-=-=-=/%
\x{SymplecticPotential[tensors][L]} computes \x{FirstVariation[tensors][L]}, discards the EOMs and strips the total derivative \x{InertHead}, returning only its argument (see \cref{eq: dlLagrangian split xCPS}).

\x{SymplecticCurrent[tensors][L]} applies \x{VertDiff} to \x{SymplecticPotential[tensors][L]} followed by \x{ExpandVertDiff} with the option \x{NonConstantTensors}$\to$\x{tensors}. The latter ensures that only dynamical fields are varied. The option \x{HoldExpandVertDiff$\to$tensor} can be added to prevent the expansion of \x{tensor} in the second variation (see the example of \cref{subsection: EM}).

\begin{mmaIn}
    SymplecticPotential[phi][LScl] // ContractMetric // Simplification\newline
SymplecticCurrent[phi][LScl] // Simplification\newline
 SymplecticCurrent[][LScl] // ContractMetric // Simplification
\end{mmaIn}
\begin{mmaOut}
    $\sqrt{-\dbtilde{g}}(\dd \phi) (n_{{}_D})^{a} D_{a}\phi$ \\
$- \sqrt{-\dbtilde{g}}(n_{{}_D})^{a} (\dd \phi) \wwedge D_{a}(\dd \phi)$\\
$- \tfrac{1}{2} \sqrt{-\dbtilde{g}}(n_{{}_D})^{a}  \Big(2 (\dd \phi) \wwedge D_{a}(\dd \phi) + (\dd \phi) \wwedge (\dd g)^{b}{}_{b} D_{a}\phi - 2 (\dd \phi) \wwedge (\dd g)_{ab} D^{b}\phi\Big)$
\end{mmaOut}
\begin{remark}
    Through considerable effort, \xPert could yield the first variation. However, it cannot compute the symplectic current since the super-commutativity of \x{WWedge} is essential.
\end{remark}

%/=-=-=-=-=-=-=\%
\subsection{Noether symmetries, potentials, and currents}\label{subsec: noether}
%\=-=-=-=-=-=-=/%
%/=*=\%
\subsubsection{NoetherSymmetryQ}
%\=*=/%
\x{NoetherSymmetryQ} checks whether a VVF is a $\d$-symmetry of a Lagrangian. Following \cref{eq: L_XL=div s}, it checks whether $\LL_\XX\mathcal{L}$ is a divergence. For that, \xCPS runs \x{DivergenceQ} (see \cref{subsubsection: divergenceQ}) over \x{VertInt[VVF][VertDiff@Lagrangian]//ExpandVertInt[]} (see \cref{subsection: Vertical operators,section: ExpandVertInt}). For example, we can use this command to check that the diffeomorphism-induced VVF, $\XX_\xi$, acting only on $\phi$ is not a $\d$-symmetry of \x{LScl} in general:

\begin{mmaIn}
vvf1 = VVFFromLieD[xi][phi]\newline
NoetherSymmetryQ[vvf1][g][LScl]\newline
NoetherSymmetryQ[vvf1][g,CheckZero$\to$True][LScl]
\end{mmaIn} 
\begin{mmaOut}
    $(\frac{\delta}{\delta \phi}) \mathcal{L}_\xi \phi$\\
    $D_{a}\phi D_{b}D^{b}\phi=0\, \&\&\, D^{c}\phi (D^{a}\xi_{c} D^{b}\phi + D^{a}\phi D^{b}\xi_{c} -  g^{ab} D_{d}\xi_{c} D^{d}\phi) + \xi^{c} (D^{b}\phi D_{c}D^{a}\phi + D^{a}\phi D_{c}D^{b}\phi -  g^{ab} D_{d}D_{c}\phi D^{d}\phi)=0\, \&\&\, \xi^{a} D_{a}D_{b}D^{b}\phi + D_{a}\xi^{a} D_{b}D^{b}\phi=D^{a}\phi D_{b}D^{b}\xi_{a} + \xi^{a} D_{b}D^{b}D_{a}\phi + 2 D_{b}D_{a}\phi D^{b}\xi^{a}$\\
    False
\end{mmaOut}
 However, if it acts on all fields, it is a symmetry:
\begin{mmaIn}
vvf2 = VVFFromLieD[xi][phi,g]\newline
NoetherSymmetryQ[vvf2][g][LScl]\newline
NoetherSymmetryQ[vvf2][g][LScl] // SortCovDs // Simplification
\end{mmaIn} 
\begin{mmaOut}
    $(\frac{\delta}{\delta g})^{ab} \mathcal{L}_\xi g_{ab} + (\frac{\delta}{\delta \phi}) \mathcal{L}_\xi \phi $\\
    $D_{a}D_{b}\xi^{b} D^{a}\phi + \xi^{a} D_{b}D^{b}D_{a}\phi = \xi^{a} D_{a}D_{b}D^{b}\phi + D^{a}\phi D_{b}D_{a}\xi^{b}$\\
    True
\end{mmaOut}
If $\xi$ is a $g$-Killing vector field, then $\mathcal{L}_\xi g_{ab}=0$. In that case \x{vvf1=vvf2} and both are symmetries.
 
%/=*=\%
\subsubsection{SymmetryPotential}
%\=*=/%
\x{SymmetryPotential[vvf][metric][L]} attempts to find the symmetry potential $\vec{s}_{\x{vvf}}$ satisfying $\LL_{\x{vvf}}\x{L}=\mathrm{div}(\vec{s}_{\x{vvf}})$ (see \cref{eq: L_XL=div s}). For that, \xCPS runs \x{FindPotentialGradient} (see \cref{subsubsection: FindPotentialGradient}) over \x{VertInt[VVF][VertDiff@Lagrangian]//ExpandVertInt[]} (see \cref{subsection: Vertical operators,section: ExpandVertInt}).
\begin{mmaIn}
SymmetryPotential[vvf1][g][LScl]
\end{mmaIn}
\begin{mmaOut}[empty]
    \mmaErr{FindPotentialGradient}{\hspace{-1ex}Loop detected at iteration 2.\hspace{-1ex} Search aborted.\hspace{-1ex} Provide the optional 'iteration' argument to obtain partial results or try applying intermediate simplification functions.}\\
    0
\end{mmaOut}
Since \x{vvf1} is not a symmetry, so no potential exists. Partial results are nonetheless available:
\begin{mmaIn}
SymmetryPotential[vvf1][g,1][LScl]\newline
SymmetryPotential[vvf1][g,2][LScl]\newline
SymmetryPotential[vvf1][g,3][LScl]
\end{mmaIn}
\begin{mmaOut}
$\tfrac{1}{2}\sqrt{-\dbtilde{g}}(n_{{}D})_{a} \xi^{a} D_{b}\phi D^{b}\phi $\\
$\tfrac{1}{2}\sqrt{-\dbtilde{g}}(n_{{}D})_{a} (2 \phi D^{a}\xi_{b} + \xi^{a} D_{b}\phi) D^{b}\phi$\vspace{.21ex}{}
\mmaErr{FindPotentialGradient}{Loop detected at iteration 2.\hspace{-1ex}  Returning the accumulated potential.\hspace{-1ex}  Try applying intermediate simplification functions.}\\
$\tfrac{1}{2}\sqrt{-\dbtilde{g}}(n_{{}D})_{a} \xi^{a} D_{b}\phi D^{b}\phi $
\end{mmaOut}

\x{vvf2}, on the other hand, does have a symmetry potential:
\begin{mmaIn}
SymmetryPotential[vvf2][g][LScl]
\end{mmaIn} 
\begin{mmaOut}
    $\tfrac{1}{2} \sqrt{-\dbtilde{g}}(n_{{}_D})_{a} \xi^{a} D_{b}\phi D^{b}\phi $
\end{mmaOut}

%/=*=\%
\subsubsection{NoetherCurrent}
%\=*=/%
\x{NoetherCurrent} attempts to extract the Noether current of a Lagrangian for a given VVF symmetry. \xCPS implements the formula given by \cref{eq: Noether current density} using \x{SymmetryPotential} (which uses in turn \x{FindPotentialGradient}, see \cref{subsubsection: FindPotentialGradient}) and \x{VertInt[vvf]} applied to \x{SymplecticPotential} (see \cref{section: ExpandVertInt,subsection: symplectic potential and current}).

Recall that \x{vvf1} is not a symmetry, so \x{NoetherCurrent} would detect a loop for the same reason as \x{SymmetryPotential} above. For \x{vvf2} we have the Noether current:
\begin{mmaIn}
NoetherCurrent[vvf2][g][LScl] // ContractMetric // Simplification%\newline
%NoetherCurrent[vvf2][g][LScl] // LieDToCovD[\#,LCDer] \& //  ContractMetric\\// Simplification
\end{mmaIn} 
\begin{mmaOut}
    $\tfrac{1}{2} \sqrt{-\dbtilde{g}}(n_{{}_D})^{a} (\xi_{a} D_{b}\phi D^{b}\phi - 2 D_{a}\phi \mathcal{L}_\xi \phi)$%\\
%    $\tfrac{1}{2}\sqrt{-\dbtilde{g}}(n_{{}_D})^{a} D_{b}\phi (-2 \xi^{b} D_{a}\phi + \xi_{a} D^{b}\phi)$
\end{mmaOut}
This expression equals $(n_{{}_D})_a\xi_b T^{ab}_{\mathrm{scl}}$ up to a factor of $-\sqrt{-g}$:
\begin{mmaIn}
-$\sqrt{\x{-Detg[]}}$NormalOfLCDer[-a]xi[-b]EnergyMomentum[LScl][a,b] \\// ContractMetric // Simplification
\end{mmaIn} 
\begin{mmaOut}
    $\tfrac{1}{2}\sqrt{-\dbtilde{g}}(n_{{}_D})^{a} D_{b}\phi ( \xi_{a} D^{b}\phi-2 \xi^{b} D_{a}\phi)$
\end{mmaOut}

%/=*=\%
\subsubsection{Non-Noether current}
%\=*=/%
\x{CurrentFromVector} computes the Non-Noether $\xi$-current for some vector field $\xi^a$. \xCPS implements the formula given by \cref{eq: Noether current density}. Unlike the Noether current, this quantity is defined for any $\xi\in\mathfrak{X}(\x{M})$, regardless of whether 
$\XX_\xi$ is a symmetry (see \cref{subsection: Non-Noether}).
\begin{mmaIn}
CurrentFromVector[xi][phi,LCDer][LScl] // ContractMetric // Simplification
\end{mmaIn} 
\begin{mmaOut}
    $\tfrac{1}{2}\sqrt{-\dbtilde{g}}(n_{{}D})^{a} (\xi_{a} D_{b}\phi D^{b}\phi - 2 D_{a}\phi \mathcal{L}_\xi \phi)$
\end{mmaOut}
Since \x{phi} is the only dynamical field, $\XX_\xi$ (i.e., \x{vvf1}) is not a symmetry and the Non-Noether charge is not conserved.

\begin{remark}
     If the metric is also taken as dynamical, then $\XX_\xi$ is \x{vvf2} and \x{CurrentFromVector} returns the same result as \x{NoetherCurrent}. The same holds if the metric is non-dynamical but $\xi$ is a $g$-Killing vector field, since $\mathcal{L}_\xi g_{ab}=0$ again makes \x{vvf1=vvf2}.
\end{remark}

%%%%%%%%%%%%%%%%%%%%%%%%%%%%%%%%%%%%%%%%%%%%%%%%%%%%%%%%%%%%%%%%%%%%%%%%%%%%%%%%%%%%%%%%%%
%%%%%%%%%%%%%%%%%%%%%%%%%%%%%%%%%%%%%%%%%%%%%%%%%%%%%%%%%%%%%%%%%%%%%%%%%%%%%%%%%%%%%%%%%%
%%%%%%%%%%%%%%%%%%%%%%%%%%%%%%%%%%%%%%%%%%%%%%%%%%%%%%%%%%%%%%%%%%%%%%%%%%%%%%%%%%%%%%%%%%

%/=====================\%
\section{Full examples}\label{sec: final examples}
%\=====================/%

In this section we introduce several detailed examples. Most functionalities require no additional knowledge beyond \xAct itself. We cover the scalar field, General Relativity, electromagnetism, and a generic $f(\mathrm{Riem})$ theory.

%/=-=-=-=-=-=-=\%
\subsection{Scalar field and General Relativity}\label{subsection: sf and GR}
%\=-=-=-=-=-=-=/%
We begin with the Klein--Gordon scalar field and General Relativity, two standard theories that illustrate the package's core functionality. For that, we begin a new session of \mathematica, load \xCPS, and define a manifold, a Lorentzian metric, and a scalar field:

\setcounter{mmaCounter}{1}
\begin{mmaIn}
<{}<xAct\`{}xCPS\`{}\newline
SetOptions[ContractMetric,AllowUpperDerivatives$\to$True];\newline
DefManifold[M,4,\{a,b,c,d,e,f,h,i,j,k,l\}]\newline
DefMetric[-1,g[-a,-b],LCDer,\{";","D"\}]\newline
DefTensor[phi[],M,PrintAs$\to$"$\phi$"]
\end{mmaIn} 
{}\vspace{-5.25ex}
%/=*=\%
\subsubsection*{First variation}
%\=*=/% 
{}\vspace{-.6ex}
We consider the first variation of the Klein--Gordon and Einstein--Hilbert Lagrangians:
\begin{mmaIn}
LScl = $\tfrac{1}{2}\sqrt{\x{-Detg[]}}$  LCDer[-a]@phi[] LCDer[a]@phi[] \newline
LGR  = $\sqrt{\x{-Detg[]}}$RicciScalarLCDer[]\newline
FirstVariation[phi][LScl]\newline
FirstVariation[][LGR] // Simplification
\end{mmaIn} 
\begin{mmaOut}
$\frac{1}{2}\sqrt{-\dbtilde{g}}\,\,g^{ab}(D_a\phi)(D_b\phi)$\\
$\sqrt{-\tilde{g}}\,R[D]$\\
$- \sqrt{-\tilde{g}} (\dd \phi) D^{b}D_{b}\phi+\x{TotalDerivativeOfLCDer}\bigl[\sqrt{-\tilde{g}} (\dd \phi) (n_{{}_D})^{a} D_{a}\phi \bigr] $\\
$\tfrac{1}{2} \sqrt{-\tilde{g}} (\dd g)^{ab} (g_{ab} R[D]\!-\!2 R[D]_{ab})+\x{TotalDerivativeOfLCDer}\bigl[\sqrt{-\tilde{g}} (n_{{}_D})^{a} (D_{b}(\dd g)_{a}{}^{b}\!-\!D_{a}(\dd g)^{b}{}_{b})\bigr]$
\end{mmaOut}
To use the inverse of the metric as the fundamental variable:
\begin{mmaIn}
VertDiff[g][-a,-b]\newline
\$UseInverseMetric = True;\mmaSup\newline\newline
VertDiff[g][-a,-b]\newline
FirstVariation[][LGR] // Simplification\newline
\$UseInverseMetric = False;\mmaSup
\end{mmaIn} 
\begin{mmaOut}
$(\dd g)_{ab}$\\
$-(\dd g^{-1})_{ab}$\\
$\tfrac{1}{2} \sqrt{-\tilde{g}} (\dd g^{-1})^{ab} (g_{ab} R[D]\!-\!2 R[D]_{ab})$\newline \hspace{10ex}${}+\x{TotalDerivativeOfLCDer}\bigl[\sqrt{-\tilde{g}} (n_{{}_D})^{a} (D_{a}(\dd g^{-1})^{b}{}_{b})\!-\!D_{b}(\dd g^{-1})_{a}{}^{b}\bigr]$
\end{mmaOut}\mmaStep

{}\vspace{-1.25ex}
%/=*=\%
\subsubsection*{Equations of motion}
%\=*=/% 
{}\vspace{-.6ex}
\begin{mmaIn}
WaveEq     = EOM[phi][LScl] // ContractMetric\newline
EinsteinEq = EOM[g][LGR] // ContractMetric // Simplification
\end{mmaIn}
\begin{mmaOut}
$- \sqrt{-\dbtilde{g}}\,\, D_a D^a \phi$\\
$\frac{1}{2}\sqrt{-\dbtilde{g}}\,\left( -2\, R[D]^{ab} + g^{ab} R[D] \right)$
\end{mmaOut}

{}\vspace{-1.25ex}
%/=*=\%
\subsubsection*{Energy--momentum tensor of the scalar field}
%\=*=/% 
{}\vspace{-.6ex}
\begin{mmaIn}
EnMomScl = EnergyMomentum[LScl] // ContractMetric // Simplification
\end{mmaIn} 
\begin{mmaOut}
$(D^{a}\phi)(D^{b}\phi)
- \frac{1}{2} g^{ab}\, (D_{c}\phi)(D^{c}\phi)$
\end{mmaOut}

{}\vspace{-1.25ex}
%/=*=\%
\subsubsection*{Wald entropy of General Relativity}
%\=*=/% 
{}\vspace{-.6ex}
\begin{mmaIn}
EntropyGR = EOM[RiemannLCDer][LGR] // Simplification
\end{mmaIn} 
\begin{mmaOut}
$\tfrac{1}{2}\sqrt{-\dbtilde{g}}(- g^{ad} g^{bc} + g^{ac} g^{bd})$
\end{mmaOut}

{}\vspace{-1.25ex}
%/=*=\%
\subsubsection*{Symplectic currents}
%\=*=/%
{}\vspace{-.6ex}
\begin{mmaIn}
$\Omega$Scl = SymplecticCurrent[phi][LScl] // ContractMetric // Simplification\newline
$\Omega$GR = SymplecticCurrent[][LGR] // ContractMetric // Simplification
\end{mmaIn}
\begin{mmaOut}
$- \sqrt{-\dbtilde{g}}(n_{{}_D})^{a} (\dd\phi) \wwedge D_{a}(\dd\phi)$\\
$\tfrac{1}{2} \sqrt{-\dbtilde{g}}(n_{{}_D})^{a} \Big((\dd g)_{a}{}^{b} \wwedge D_{b}(\dd g)^{c}{}_{c} -  (\dd g)^{b}{}_{b} \wwedge D_{a}(\dd g)^{c}{}_{c} + (\dd g)^{b}{}_{b} \wwedge D_{c}(\dd g)_{a}{}^{c}$\newline$ +{}\,(\dd g)^{bc} \wwedge D_{a}(\dd g)_{bc} - 2 (\dd g)^{bc} \wwedge D_{c}(\dd g)_{ab}\Big)$
\end{mmaOut}
This coincides, as expected, with the theoretical results \cite{barbero2021covariant}.

{}\vspace{-1.25ex}
%/=*=\%
\subsubsection*{Noether symmetries}
%\=*=/%
{}\vspace{-.6ex}

\begin{mmaIn}
DefTensor[xi[a],M,PrintAs$\to$"$\xi$"]\mmaSup\newline
vvf1 = VVFFromLieD[xi][phi]\newline
vvf2 = VVFFromLieD[xi][phi,g]\newline
NoetherSymmetryQ[vvf1][g,CheckZero$\to$True][LScl] \newline
NoetherSymmetryQ[vvf2][g][LScl] // SortCovDs // Simplification
\end{mmaIn}
\begin{mmaOut}
$ (\frac{\delta}{\delta \phi}) \mathcal{L}_\xi \phi $\\
$(\frac{\delta}{\delta g})^{ab} \mathcal{L}_\xi g_{ab} + (\frac{\delta}{\delta \phi}) \mathcal{L}_\xi \phi $\\
False\\
True
\end{mmaOut}
\begin{mmaIn}
ApplyBianchi[expr\_]:=···;\mmaSup\newline
vvf3 = VVFFromLieD[xi][g]\newline
NoetherSymmetryQ[vvf3][g][LGR] // SortCovDs // Simplification \\// 
   ApplyBianchi // ContractMetric // Simplification
\end{mmaIn}
\begin{mmaOut}
$(\frac{\delta}{\delta g})^{ab} \mathcal{L}_\xi g_{ab}$\\
True
\end{mmaOut}
\x{ApplyBianchi} is a user-defined function, provided in the companion file, that applies the Bianchi identity in cases such as those arising here. The package \TInvar implements a more systematic approach with a database of identities \cite{TInvar}.

{}\vspace{-1.25ex}
%/=*=\%
\subsubsection*{Noether currents of the scalar field}
%\=*=/%
{}\vspace{-.6ex}

\begin{mmaIn}
NoetherCurrent[vvf1][g][LScl] // ContractMetric // Simplification\newline
NoetherCurrent[vvf2][g][LScl] // ContractMetric // Simplification
\end{mmaIn}
\begin{mmaOut}[empty]
    \mmaErr{FindPotentialGradient}{\hspace{-1ex}Loop detected at iteration 2.\hspace{-1ex} Search aborted.\hspace{-1ex} Provide the optional 'iteration' argument to obtain partial results or try applying intermediate simplification functions.}\\
    $-\sqrt{-\dbtilde{g}}(n_{{}D})^{a} D_{a}\phi \mathcal{L}_\xi \phi $\\
$\tfrac{1}{2}\sqrt{-\dbtilde{g}}(n_{{}D})^{a} (\xi_{a} D_{b}\phi D^{b}\phi - 2 D_{a}\phi \mathcal{L}_\xi \phi)$
\end{mmaOut}
\x{vvf1} is not a symmetry but \x{vvf2} is; see \cref{subsec: noether} for the detailed explanation.

{}\vspace{-1.25ex}
%/=*=\%
\subsubsection*{Non-Noether currents of the scalar field}
%\=*=/%
{}\vspace{-.6ex}

\begin{mmaIn}
CurrentFromVector[xi][g][LScl] // LieDToCovD[\#,LCDer] \& // ContractMetric\\ // Simplification
\end{mmaIn}
\begin{mmaOut}
$\tfrac{1}{2} \sqrt{-\dbtilde{g}}(n_{{}_D})^{a} D_{b}\phi (-2 \xi^{b} D_{a}\phi + \xi_{a} D^{b}\phi)$
\end{mmaOut}
Integrating over a Cauchy surface $\Sigma$ yields the $\xi$-energy of $\Sigma$. It is not conserved in general: evaluating on a different Cauchy surface $\Sigma'$ yields a different value even on-shell.

{}\vspace{-1.25ex}
%/=*=\%
\subsubsection*{Noether currents of GR}
%\=*=/%
{}\vspace{-.6ex}

\begin{mmaIn}
NoetherCurrent[vvf3][g][LGR]
\end{mmaIn}
\begin{mmaOut}[empty]
    \mmaErr{FindPotentialGradient}{\hspace{-1ex}Loop detected at iteration 4.\hspace{-1ex} Search aborted.\hspace{-1ex} Provide the optional 'iteration' argument to obtain partial results or try applying intermediate simplification functions.}\\
    $\sqrt{-\dbtilde{g}}\, g^{bc} (n_{{}D})^{a} (D_{a}\mathcal{L}_\xi g_{bc} -  D_{c}\mathcal{L}_\xi g_{ab})$
\end{mmaOut}
\x{NoetherCurrent} finds a loop even though \x{vvf3} is a symmetry of \x{LGR}. The problem arises because \x{SymmetryPotential} reorders covariant derivatives without applying the Bianchi identities, causing the algorithm to fail to recognize that intermediate expressions vanish. We can force the application of the Bianchi identities in the intermediate steps as follows:
\begin{mmaIn}
SymmetryPotential[vvf3][g][LGR,SortCovDs,ApplyBianchi]\newline
NoetherCurrent[vvf3][g][LGR,ApplyBianchi] // LieDToCovD[\#,LCDer] \& \\//
    SortCovDs // ContractMetric //
 Simplification
\end{mmaIn}
\begin{mmaOut}
$\sqrt{-\dbtilde{g}}(n_{{}_D})_{a} R[D] \xi^{a}$\\
$\sqrt{-\dbtilde{g}}(n_{{}_D})^{a} (R[D] \xi_{a} - 2 R[D]_{ab} \xi^{b})$
\end{mmaOut}
The Noether current vanishes on-shell, as expected.

\begin{remark}
Different choices of intermediate functions can affect the form of the result. When they do, the outputs differ by a total divergence and are thus equivalent. For instance:
\begin{mmaIn}
NoetherCurrent[vvf3][g][LGR,ApplyBianchi,SortCovDs] // LieDToCovD[\#,LCDer] \& \\//
    SortCovDs // ContractMetric //
 Simplification
\end{mmaIn}
\begin{mmaOut}
$\sqrt{-\dbtilde{g}}(n_{{}_D})^{a} (R[D] \xi_{a} - 2 R[D]_{ab} \xi^{b} + D_{b}D_{a}\xi^{b} -  D_{b}D^{b}\xi_{a})$
\end{mmaOut}
\end{remark}

%/=-=-=-=-=-=-=\%
\subsection{Electromagnetism}\label{subsection: EM}
%\=-=-=-=-=-=-=/%
We now treat electromagnetism (EM). We begin a new \mathematica session and define a manifold, a Lorentzian metric, a 1-form $A_a$, and an antisymmetric tensor $F_{ab}$:

\setcounter{mmaCounter}{1}
\begin{mmaIn}
<{}<xAct\`{}xCPS\`{}\newline
SetOptions[ContractMetric,AllowUpperDerivatives$\to$True];\newline
DefManifold[M,4,\{a,b,c,d,e,f,h,i,j,k,l\}]\newline
DefMetric[-1,g[-a,-b],LCDer,\{";","D"\}]\newline
DefTensor[A[-a],M]\newline
DefTensor[F[-a,-b],M,Antisymmetric[{-a,-b}]]
\end{mmaIn} 
 
%/=*=\%
\subsubsection*{First variation}
%\=*=/% 
{}\vspace{-.6ex}
We define the EM Lagrangian and include the expansion rule for $\dd F$:
\begin{mmaIn}
LEM = $\tfrac{1}{4} \sqrt{\x{-Detg[]}}$ F[-a,-b] F[a,b]\newline
GenerateExpandVertDiffRule[{dlF[-a,-b],LCDer[-a]@dlA[-b]-LCDer[-b]@dlA[-a]}]\mmaSup\newline
VertDiff[F[-a,-b]] // ExpandVertDiff[]
\end{mmaIn} 
\begin{mmaOut}
$\tfrac{1}{4}\sqrt{-\dbtilde{g}}\,F_{ab} F^{ab}$\\
$D_{a}(\dd A)_{b} -  D_{b}(\dd A)_{a}$
\end{mmaOut}

Its first variation is given by:
\begin{mmaIn}
FirstVariation[A,LCDer][LEM]
\end{mmaIn} 
\begin{mmaOut}
$\sqrt{-\dbtilde{g}}(\dd A)_{a} D_{b}F^{ab}+\x{TotalDerivativeOfLCDer}\bigl[- \sqrt{-\dbtilde{g}}(\dd A)^{a} F_{ab} (n_{{}_D})^{b}\bigr]$
\end{mmaOut}
We can see that the Gauss law is written in a compact form in terms of the $F$ tensor. This is one of the novelties of \xCPS.

{}\vspace{-1.25ex}
%/=*=\%
\subsubsection*{Energy-momentum tensor}
%\=*=/%
{}\vspace{-.6ex}
\begin{mmaIn}
EnMomEM = EnergyMomentum[LEM] // ContractMetric // Simplification
\end{mmaIn}
\begin{mmaOut}
$F^{ac} F^{b}{}_{c} -  \tfrac{1}{4} F_{cd} F^{cd} g^{ab}$
\end{mmaOut}

{}\vspace{-1.25ex}
%/=*=\%
\subsubsection*{Symplectic currents}
%\=*=/%
{}\vspace{-.6ex}
\begin{mmaIn}
SymplecticCurrent[A,LCDer][LEM] // ContractMetric // Simplification\newline
SymplecticCurrent[A,LCDer,HoldExpandVertDiff$\to$F][LEM] // ContractMetric\\ // Simplification
\end{mmaIn}
\begin{mmaOut}
$- \sqrt{-\dbtilde{g}}(n_{{}_D})^{a} ( (\dd A)^{b} \wwedge D_{b}(\dd A)_{a}- (\dd A)^{b} \wwedge D_{a}(\dd A)_{b} )$\\
$- \sqrt{-\dbtilde{g}}(n_{{}_D})^{a} (\dd A)^{b} \wwedge (\dd F)_{ab}$
\end{mmaOut}
The second form is more compact since \x{HoldExpandVertDiff} prevents $\dd F$ from being expanded into $\dd A$ terms in the second variation, keeping the field strength manifest.

{}\vspace{-1.25ex}
%/=*=\%
\subsubsection*{Noether symmetries}
%\=*=/%
{}\vspace{-.6ex}
\begin{mmaIn}
DefTensor[lambda[],M,PrintAs$\to$"$\lambda$"]\mmaSup\newline
DefTensor[xi[a],M,PrintAs$\to$"$\xi$"]\mmaSup\newline
vvfEM1 = LCDer[-a]@lambda[] VariationalVectorA[a]\newline
vvfEM2 = VVFFromLieD[xi][A]\newline
NoetherSymmetryQ[vvfEM1][g][LEM]\newline
NoetherSymmetryQ[vvfEM2][g,CheckZero$\to$True][LEM]
\end{mmaIn}
\begin{mmaOut}
$(\frac{\delta}{\delta A})^{a} D_{a}\lambda $\\
$(\frac{\delta}{\delta A})^{a} \mathcal{L}_\xi A_{a}$\\
True\\
False
\end{mmaOut}
\x{vvfEM1} corresponds to the standard EM gauge transformation.

{}\vspace{-1.25ex}
%/=*=\%
\subsubsection*{Noether current}
%\=*=/%
{}\vspace{-.6ex}
\begin{mmaIn}
NoetherCurrent[vvfEM1][LCDer][LEM] // ContractMetric // Simplification
\end{mmaIn}
\begin{mmaOut}
$- \sqrt{-\dbtilde{g}}\,F_{ab} (n_{{}_D})^{a} D^{b}\lambda $
\end{mmaOut}
Upon integration over a closed Cauchy surface and integration by parts, this expression vanishes on-shell: the extrinsic curvature term that arises cancels by the antisymmetry of $F$.

{}\vspace{-1.25ex}
%/=*=\%
\subsubsection*{Non-Noether current}
%\=*=/%
{}\vspace{-.6ex}
\begin{mmaIn}
CurrentFromVector[xi][A,LCDer][LEM] // LieDToCovD[\#,LCDer] \& \\// 
  ContractMetric // Simplification
\end{mmaIn}
\begin{mmaOut}
$ \sqrt{-\dbtilde{g}}(n_{{}_D})^{a}\Big( \tfrac{1}{4} F_{bc} F^{bc} \xi_{a} -  F_{ac} \xi^{b} D_{b}A^{c} -  A^{c} F_{ab} D^{b}\xi_{c}\Big)$
\end{mmaOut}
This is the EM energy-momentum current. It is not conserved in general since $\XX_\xi$ is not a symmetry ($g$ is non-dynamical). Conservation is restored if, for instance, $\xi$ is a $g$-Killing vector field.

%/=-=-=-=-=-=-=\%
\subsection{f(Riem)}
%\=-=-=-=-=-=-=/%
We now consider a generic $f(\x{Riem})$ theory. As before, we begin a new \mathematica session and define a manifold and a Lorentzian metric:
\setcounter{mmaCounter}{1}
\begin{mmaIn}
<{}<xAct\`{}xCPS\`{}\newline
SetOptions[ContractMetric,AllowUpperDerivatives$\to$True];\newline
DefManifold[M,4,\{a,b,c,d,e,f,h,i,j,k,l\}]\newline
DefMetric[-1,g[-a,-b],LCDer,\{";","D"\}]\newline
PrintAs[RiemannLCDer] $\verb|^=|$ "Riem";
\end{mmaIn} 
To define the Lagrangian, we use the \xAct command \x{DefScalarFunction}, which is modified by \xCPS to accept an additional argument: the list of tensors on which it depends variationally.
\begin{mmaIn}
DefScalarFunction[L,{RiemannLCDer}]\mmaSup\newline
Ldensity = $\sqrt{\x{-Detg[]}}$ L[RiemannLCDer]
\end{mmaIn} 
\begin{mmaOut}
$\sqrt{-\tilde{\tilde{g}}}\, L\bigl[Riem\bigr]$
\end{mmaOut}

{}\vspace{-1.25ex}
%/=*=\%
\subsubsection*{EOM}
%\=*=/%
{}\vspace{-.6ex}

\begin{mmaIn}
EOM[g][Ldensity] // ContractMetric // Simplification
  \end{mmaIn}
\begin{mmaOut}
$\tfrac{1}{2}\sqrt{-\tilde{\tilde{g}}} \Bigl(g^{ab} L\bigl[Riem\bigr] + (\frac{\partial L}{\partial Riem})^{bcde} Riem^{a}{}_{cde} + (\frac{\partial L}{\partial Riem})^{acde} Riem^{b}{}_{cde}$ \newline $\hspace{10ex}- 2 D_{c}D_{d}(\frac{\partial L}{\partial Riem})^{acbd} - 2 D_{d}D_{c}(\frac{\partial L}{\partial Riem})^{acbd}\Bigr)$
\end{mmaOut}
\x{ExpandVertDiff} uses the chain rule when acting over the \x{VertDiff} of a scalar function. If not yet defined, the command defines the required tensors $\frac{\partial\x{function}}{\partial \x{tensor}}$ automatically. The indices are opposite to \x{tensor} and with the same symmetries. Its weight equals the weight of \x{function} minus that of \x{tensor}. These tensors are called \x{PartialPartial} tensors within \xCPS.

{}\vspace{-1.25ex}
%/=*=\%
\subsubsection*{Wald entropy}
%\=*=/%
{}\vspace{-.6ex}

\begin{mmaIn}
EOM[RiemannLCDer][Ldensity] // Simplification
  \end{mmaIn}
\begin{mmaOut}
$\sqrt{-\tilde{\tilde{g}}} (\frac{\partial L}{\partial Riem})^{abcd}$
\end{mmaOut}

{}\vspace{-1.25ex}
%/=*=\%
\subsubsection*{Symplectic Current}
%\=*=/%
{}\vspace{-.6ex}

\begin{mmaIn}
SymplecticCurrent[][Ldensity] // ContractMetric // Simplification
  \end{mmaIn} 
\begin{mmaOut}
$\sqrt{-\tilde{\tilde{g}}}  (n_{{}_D})^{a} \biggl((\frac{\partial L}{\partial Riem})^{bcde} (\dd g)_{bd} \wwedge D_{a}(\dd g)_{ce} - 2 (\frac{\partial L}{\partial Riem})^{bcde} (\dd g)_{bd} \wwedge D_{e}(\dd g)_{ac}$ \newline
$+ 2  (\frac{\partial^{2} L}{\partial^{2}Riem})_{a}{}^{fhi}{}_{bcd}{}^{j} Riem^{bcde} (\dd g)_{ej} \wwedge D_{i}(\dd g)_{fh} + (\frac{\partial L}{\partial Riem})_{a}{}^{bcd} \big\{(\dd g)_{bc} \wwedge D_{d}(\dd g)^{e}{}_{e}  $\newline
$+2 (\dd g)_{c}{}^{e} \wwedge D_{b}(\dd g)_{de} + 2 (\dd g)_{c}{}^{e} \wwedge D_{d}(\dd g)_{be} - 2 (\dd g)_{c}{}^{e} \wwedge D_{e}(\dd g)_{bd} + (\dd g)^{e}{}_{e} \wwedge D_{d}(\dd g)_{bc}\bigr\}$\newline
$ + 2 (\frac{\partial^{2} L}{\partial^{2}Riem})_{a}{}^{fhi}{}_{bcd}{}^{j} Riem^{bcde} (\dd g)_{fh} \wwedge D_{i}(\dd g)_{ej} - 4 (\frac{\partial^{2} L}{\partial^{2}Riem})_{a}{}^{bcdefhi} \bigl\{(\dd g)_{bc} \wwedge D_{d}D_{i}D_{f}(\dd g)_{eh} $\newline
$-  D_{d}(\dd g)_{bc} \wwedge D_{i}D_{f}(\dd g)_{eh}\bigr\} + 4 (\dd g)_{bd} \wwedge D_{i}D_{f}(\dd g)_{eh} D_{c}(\frac{\partial^{2} L}{\partial^{2}Riem})_{a}{}^{bcdefhi} $\newline
$-  (\dd g)_{bd} \wwedge (\dd g)^{e}{}_{e} D_{c}(\frac{\partial L}{\partial Riem})_{a}{}^{bcd} + 2 (\frac{\partial^{2} L}{\partial^{2}Riem})_{a}{}^{h}{}_{f}{}^{i}{}_{bcd}{}^{j} (\dd g)_{ej} \wwedge (\dd g)_{hi} D^{f}Riem^{bcde} $\newline
$+ 2 Riem^{bcde} (\dd g)_{ej} \wwedge (\dd g)_{fi} D_{h}(\frac{\partial^{2} L}{\partial^{2}Riem})_{a}{}^{fhi}{}_{bcd}{}^{j}\biggr)$
\end{mmaOut}
Higher-order derivatives of $L$ are handled analogously: \x{ExpandVertDiff} applies the chain rule and defines (if needed) the required higher-order \x{PartialPartial} tensors of the form
\[\frac{\partial^k\x{function}}{\partial^{i_1} \x{tensor}_1\cdots{}\partial^{i_m} \x{tensor}_m}\]
The indices are the concatenation of the opposite indices of $\x{tensor}_1,\ldots,\x{tensor}_m$ (the tensors are always ordered lexicographically, since the ``partial derivatives'' commute by assumption). The symmetry group of this tensor is nontrivial:
\begin{enumerate}
    \item It inherits the symmetries of each $\x{tensor}_i$ for the corresponding indices.
    \item When the same tensor appears more than once, the object is symmetric under exchange of the corresponding index groups:
    \begin{mmaIn}
    expr = Partial2LPartial2RiemannLCDer[a,b,c,d,e,f,h,i] - \\
     Partial2LPartial2RiemannLCDer[e,f,h,i,a,b,c,d]\newline
    expr // ToCanonical\newline
      \end{mmaIn} 
    \begin{mmaOut}
    $(\frac{\partial^{2} L}{\partial^{2}Riem})^{abcdefhi} -  (\frac{\partial^{2} L}{\partial^{2}Riem})^{efhiabcd}$\\
    $0$
    \end{mmaOut}
\end{enumerate}
\x{PartialPartial} tensors are one of the main innovations introduced in \xCPS, enabling, to the best of our knowledge for the first time, the treatment of generic Lagrangians.

{}\vspace{-1.25ex}
%/=*=\%
\subsubsection*{Noether symmetries}
%\=*=/%
{}\vspace{-.6ex}

Noether symmetries can be studied but the generic Lagrangian produces lengthy intermediate expressions. We display only the condition with respect to $\xi$:

\begin{mmaIn}
DefTensor[xi[a],M,PrintAs$\to$"$\xi$"]\mmaSup\newline
vvf1 = VVFFromLieD[xi][g];\mmaSup\newline
NoetherSymmetryQ[vvf1][g][Ldensity][[1]] // SortCovDs // Simplification
  \end{mmaIn} 
\begin{mmaOut}
$Riem^{bcde} D_{c}(\frac{\partial L}{\partial Riem})_{abde} + (\frac{\partial L}{\partial Riem})^{bcde} D_{c}Riem_{abde} + Riem_{a}{}^{bcd} D_{e}(\frac{\partial L}{\partial Riem})_{b}{}^{e}{}_{cd} $ \newline$+ (\frac{\partial L}{\partial Riem})_{a}{}^{bcd} D_{e}Riem_{b}{}^{e}{}_{cd} = D_{a}L\bigl[Riem\bigr] + 2 \bigl(D_{d}D_{b}D_{c}(\frac{\partial L}{\partial Riem})_{a}{}^{bcd} + D_{d}D_{c}D_{b}(\frac{\partial L}{\partial Riem})_{a}{}^{bcd}\bigr)$
\end{mmaOut}

{}\vspace{-1.25ex}
%/=*=\%
\subsubsection*{Non-Noether current}
%\=*=/%
{}\vspace{-.6ex}

\begin{mmaIn}
CurrentFromVector[xi][LCDer][Ldensity] // LieDToCovD[\#, LCDer] \& // 
  ContractMetric // Simplification
  \end{mmaIn} 
\begin{mmaOut}
$\sqrt{-\tilde{\tilde{g}}} (n_{{}_D})^{a} \Bigl(L[Riem\bigr] \xi_{a} + 2 D^{c}\xi^{b} \big\{D_{d}(\frac{\partial L}{\partial Riem})_{abc}{}^{d} + D_{d}(\frac{\partial L}{\partial Riem})_{acb}{}^{d}\big\} $\newline
$\hspace{15ex}- 2 \big\{(\frac{\partial L}{\partial Riem})_{abcd} + (\frac{\partial L}{\partial Riem})_{acbd}\big\} D^{d}D^{c}\xi^{b}\Bigr)$
\end{mmaOut}

{}\vspace{-1.25ex}
%/=*=\%
\subsubsection*{Higher order theory}
%\=*=/%
{}\vspace{-.6ex}

\xCPS also handles theories depending on derivatives of a tensor. For that we  use \x{Implode} (see \cref{subsection: quick tour xact}) to create the tensor \x{LCDerRiemannLCDer}, representing $D_a\mathrm{Riem}_{bcde}$:

\begin{mmaIn}
    Implode[LCDer[-a][RiemannLCDer[-b,-c,-d,-e]]]\newline
DefScalarFunction[L2,\{RiemannLCDer,LCDerRiemannLCDer\}]\newline
Ldensity2 = $\sqrt{\x{-Detg[]}}$ L2[RiemannLCDer,LCDerRiemannLCDer]\newline
Entropy2 = EOM[RiemannLCDer][Ldensity2] // Simplification
\end{mmaIn}
\begin{mmaOut}
    $DRiem_{bcdea}$\\
    $\sqrt{-\tilde{\tilde{g}}} L2\bigl[Riem, DRiem\bigr]$\\
    $\sqrt{-\tilde{\tilde{g}}} ((\frac{\partial L2}{\partial Riem})^{abcd} -  D_{e}(\frac{\partial L2}{\partial DRiem})^{abcde})$
\end{mmaOut}
We have computed for concreteness only the Wald entropy, which now receives a correction from the derivative coupling. All other quantities — equations of motion, symplectic current, Noether charges — can be computed in exactly the same way as for the $f(\mathrm{Riem})$ case above.

%%%%%%%%%%%%%%%%%%%%%%%%%%%%%%%%%%%%%%%%%%%%%%%%%%%%%%%%%%%%%%%%%%%%%%%%%%%%%%%%%%%%%%%%%%
%%%%%%%%%%%%%%%%%%%%%%%%%%%%%%%%%%%%%%%%%%%%%%%%%%%%%%%%%%%%%%%%%%%%%%%%%%%%%%%%%%%%%%%%%%
%%%%%%%%%%%%%%%%%%%%%%%%%%%%%%%%%%%%%%%%%%%%%%%%%%%%%%%%%%%%%%%%%%%%%%%%%%%%%%%%%%%%%%%%%%
 
%/=====================\%
\section{Conclusions}\label{sec: conclusions}
%\=====================/%

We have introduced \xCPS, a \mathematica package built upon the \xAct tensor algebra bundle, designed to automate symbolic computations within the covariant phase space formalism. Starting from a specified Lagrangian density, \xCPS systematically derives equations of motion, symplectic potentials, and both Noether and non-Noether currents. As a highly useful byproduct, it also introduces an exact divergence-detection algorithm and a heuristic potential-finding algorithm, both of which hold independent computational value beyond the phase space framework.

The package introduces several novel features that, to the best of our knowledge, remain unavailable in any prior tensor algebra implementation. Chief among these is the accommodation of generic Lagrangians via \x{PartialPartial} tensors, which encode functional derivatives of unspecified scalar functions with respect to a tensor and its derivatives. This innovation allows \xCPS to seamlessly manipulate generalized frameworks such as $f(\mathrm{Riem})$ gravity and $L(\mathrm{Riem},\nabla\mathrm{Riem})$ theories without demanding a fixed functional form. Furthermore, the variational graph and its recursive constantness-propagation algorithm provide a systematic, extensible mechanism for tracking tensor dependencies, resolving the structural limitations of the \x{VarD} approach. Finally, a native implementation of vertical exterior calculus—featuring the supercommutative \x{WWedge} product and the four foundational vertical operators—makes the symplectic current computable in absolute generality, a milestone unachievable via perturbation-based tools such as \xPert.

The applications presented in \cref{sec: final examples} demonstrate the reliability and expansive reach of the package. For standard frameworks like the Klein--Gordon scalar field, electromagnetism, and General Relativity, \xCPS reproduces known expressions. For higher-derivative gravity, the package effortlessly generates highly complex outputs that would be exceptionally tedious, if not impossible, to derive by hand or with pre-existing software.

Several development ideas remain open:
\begin{itemize}
    \item \textbf{Algorithmic Enhancements:} The divergence-detection algorithm, \x{DivergenceQ}, currently requires a scalar input; extending it to expressions with free indices would significantly broaden its applicability.
    \item \textbf{Ecosystem Integration:} A basic compatibility with \xCoba is already in place but further integration and testing is required. A combination with \xTerior is also desirable. Together, the two packages would provide a complete implementation of the variational bicomplex, handling both horizontal and vertical differential forms within a unified framework. However, this relies on a double-graded wedge product, requiring a non-trivial addition to \xAct.
    \item \textbf{Physical Extensions:} Finally, extending \xCPS to accommodate manifolds with boundaries \cite{margalef2021geometric} will unlock the automated computation of boundary charges and corner terms, which are increasingly vital in the modern study of asymptotic symmetries and edge modes.
\end{itemize}
By automating the tedious and error-prone derivations inherent to the CPS formalism, we hope \xCPS will serve as a powerful and reliable asset for the theoretical physics community.

%/=====================\%
% Bibliography
%\=====================/%
\bibliographystyle{plain}\small
\bibliography{sample.bib}

@article{margalef2018towards,
  title={Towards general relativity through parametrized theories},
  author={Margalef-Bentabol, Juan},
  journal={arXiv preprint arXiv:1807.05534},
  year={2018}
}

@incollection{zuckerman1987action,
  title={Action principles and global geometry},
  author={Zuckerman, Gregg J},
  booktitle={Mathematical aspects of string theory},
  pages={259--284},
  year={1987},
  publisher={World Scientific}
}

@incollection{Crnkovic1986,
  title     = {Covariant description of canonical formalism 
               in geometrical theories},
  author    = {Čedomir Crnković
 and Witten, Edward},
  booktitle = {Three Hundred Years of Gravitation},
  editor    = {Hawking, Stephen W. and Israel, Werner},
  pages     = {676--684},
  year      = {1987},
  publisher = {Cambridge University Press}
}

@article{margalef2021geometric,
  title={Geometric formulation of the Covariant Phase Space 
         methods with boundaries},
  author={Margalef-Bentabol, Juan and Villase{\~n}or, Eduardo JS},
  journal={\href{https://doi.org/10.1103/PhysRevD.103.025011}
           {Physical Review D}},
  volume={103},
  number={2},
  pages={025011},
  year={2021},
  publisher={APS}
}

@article{barbero2021covariant,
  title={Covariant phase space for gravity with boundaries: 
         metric versus tetrad formulations},
  author={Barbero G, J Fernando and Margalef-Bentabol, Juan 
          and Varo, Valle and Villase{\~n}or, Eduardo JS},
  journal={\href{https://doi.org/10.1103/PhysRevD.104.044048}
           {Physical Review D}},
  volume={104},
  number={4},
  pages={044048},
  year={2021},
  publisher={APS}
}

@article{barbero2021palatini,
  title={Palatini gravity with nonmetricity, torsion, and 
         boundaries in metric and connection variables},
  author={Barbero G, J Fernando and Margalef-Bentabol, Juan 
          and Varo, Valle and Villase{\~n}or, Eduardo JS},
  journal={\href{https://doi.org/10.1103/PhysRevD.104.044046}
           {Physical Review D}},
  volume={104},
  number={4},
  pages={044046},
  year={2021},
  publisher={APS}
}

@article{barbero2022shell,
  title={On-shell equivalence of general relativity and {H}olst theories with nonmetricity, torsion, and boundaries},
  author={Barbero G, J Fernando and Margalef-Bentabol, Juan 
          and Varo, Valle and Villase{\~n}or, Eduardo JS},
  journal={\href{https://doi.org/10.1103/PhysRevD.105.064066}
           {Physical Review D}},
  volume={105},
  number={6},
  pages={064066},
  year={2022},
  publisher={APS}
}

@article{margalef2022proof,
  title={Proof of the equivalence of the symplectic forms 
         derived from the canonical and the covariant phase 
         space formalisms},
  author={Margalef-Bentabol, Juan and Villase{\~n}or, Eduardo JS},
  journal={\href{https://doi.org/10.1103/PhysRevD.105.L101701}
           {Physical Review D}},
  volume={105},
  number={10},
  pages={L101701},
  year={2022},
  publisher={APS}
}

@techreport{anderson1989variational,
  title     = {The variational bicomplex},
  author    = {Anderson, Ian M},
  year      = {1989},
  institution = {Utah State University},
  note      = {Available at \url{http://math.usu.edu/~fg_mp}}
}

@article{iyer1994some,
  title={Some properties of the Noether charge and a proposal 
         for dynamical black hole entropy},
  author={Iyer, Vivek and Wald, Robert M},
  journal={\href{https://doi.org/10.1103/PhysRevD.50.846}
           {Physical Review D}},
  volume={50},
  number={2},
  pages={846},
  year={1994},
  publisher={APS}
}

@article{lee1990local,
  title={Local symmetries and constraints},
  author={Lee, Joohan and Wald, Robert M},
  journal={\href{https://doi.org/10.1063/1.528801}
           {Journal of Mathematical Physics}},
  volume={31},
  number={3},
  pages={725--743},
  year={1990},
  publisher={American Institute of Physics}
}

@article{wald1993black,
  title={Black hole entropy is the Noether charge},
  author={Wald, Robert M},
  journal={\href{https://doi.org/10.1103/PhysRevD.48.R3427}
           {Physical Review D}},
  volume={48},
  number={8},
  pages={R3427},
  year={1993},
  publisher={APS}
}

@article{wald1990identically,
  title={On identically closed forms locally constructed 
         from a field},
  author={Wald, Robert M},
  journal={\href{https://doi.org/10.1063/1.528839}
           {Journal of Mathematical Physics}},
  volume={31},
  number={10},
  pages={2378--2384},
  year={1990},
  publisher={American Institute of Physics}
}

@article{wald2000general,
  title={General definition of ``conserved quantities'' in 
         general relativity and other theories of gravity},
  author={Wald, Robert M and Zoupas, Andreas},
  journal={\href{https://doi.org/10.1103/PhysRevD.61.084027}
           {Physical Review D}},
  volume={61},
  number={8},
  pages={084027},
  year={2000},
  publisher={APS}
}

@article{bondi1962gravitational,
  title={Gravitational waves in general relativity, {VII}. 
         Waves from axi-symmetric isolated system},
  author={Bondi, Hermann and Van der Burg, MG Julian 
          and Metzner, AW Kenneth},
  journal={\href{https://doi.org/10.1098/rspa.1962.0161}
           {Proceedings of the Royal Society of London. 
           Series A. Mathematical and Physical Sciences}},
  volume={269},
  number={1336},
  pages={21--52},
  year={1962},
  publisher={The Royal Society London}
}

@article{sachs1962asymptotic,
  title={Asymptotic symmetries in gravitational theory},
  author={Sachs, Rainer K.},
  journal={\href{https://doi.org/10.1103/PhysRev.128.2851}
           {Physical Review}},
  volume={128},
  number={6},
  pages={2851--2864},
  year={1962},
  publisher={APS}
}

@article{cachazo2014evidence,
  title={Evidence for a new soft graviton theorem},
  author={Cachazo, Freddy and Strominger, Andrew},
  journal={\href{https://arxiv.org/abs/1404.4091}
           {arXiv preprint arXiv:1404.4091}},
  year={2014}
}

@article{campiglia2014asymptotic,
  title={Asymptotic symmetries and subleading soft graviton 
         theorem},
  author={Campiglia, Miguel and Laddha, Alok},
  journal={\href{https://doi.org/10.1103/PhysRevD.90.124028}
           {Physical Review D}},
  volume={90},
  number={12},
  pages={124028},
  year={2014},
  publisher={APS}
}

@article{barnich2011bms,
  title={{BMS} charge algebra},
  author={Barnich, Glenn and Troessaert, Cedric},
  journal={\href{https://doi.org/10.1007/JHEP12(2011)105}
           {Journal of High Energy Physics}},
  volume={2011},
  number={12},
  pages={1--22},
  year={2011},
  publisher={Springer}
}

@article{freidel2021weyl,
  title={The {Weyl} {BMS} group and {Einstein}'s equations},
  author={Freidel, Laurent and Oliveri, Roberto 
          and Pranzetti, Daniele and Speziale, Simone},
  journal={\href{https://doi.org/10.1007/JHEP07(2021)170}
           {Journal of High Energy Physics}},
  volume={2021},
  number={7},
  pages={170},
  year={2021},
  publisher={Springer}
}

@book{wald2010general,
  title={General relativity},
  author={Wald, Robert M},
  year={2010},
  publisher={University of Chicago Press}
}

@misc{TexAct,
  author       = {B\"ackdahl, Thomas and Mart\'{\i}n-Garc\'{\i}a, 
                  Jos\'e M. and Wardell, Barry},
  title        = {{TexAct}: {T}e{X} code to format {xAct} 
                  expressions},
  howpublished = {\href{https://github.com/xAct-contrib/TexAct}
                  {\texttt{github.com/xAct-contrib/TexAct}}},
  note         = {Mathematica package}
}

@misc{xTerior,
  author       = {Garc\'{\i}a-Parrado, Alfonso and Stein, Leo C.},
  title        = {{xTerior}: Exterior calculus in {Mathematica}},
  howpublished = {\href{https://github.com/xAct-contrib/xTerior}
                  {\texttt{github.com/xAct-contrib/xTerior}}},
  note         = {Mathematica package}
}

@article{xPert,
  author       = {Brizuela, David and Mart\'{\i}n-Garc\'{\i}a, 
                  Jos\'e M. and Mena Marug\'an, Guillermo A.},
  title        = {{xPert}: Computer algebra for metric 
                  perturbation theory},
  journal      = {\href{https://doi.org/10.1007/s10714-009-0773-2}
                  {General Relativity and Gravitation}},
  volume       = {41},
  pages        = {2415--2431},
  year         = {2009},
  archivePrefix = {arXiv},
  eprint       = {0807.0824},
  primaryClass = {gr-qc}
}

@article{xPerm,
  author       = {Mart\'{\i}n-Garc\'{\i}a, Jos\'e M.},
  title        = {{xPerm}: fast index canonicalization for 
                  tensor computer algebra},
  journal      = {\href{https://doi.org/10.1016/j.cpc.2008.05.009}
                  {Computer Physics Communications}},
  volume       = {179},
  pages        = {597--603},
  year         = {2008},
  archivePrefix = {arXiv},
  eprint       = {0803.0862},
  primaryClass = {cs.SC}
}

@misc{xCPS,
  author       = {Margalef-Bentabol, Juan and 
                  S\'anchez Cotta, Laura},
  title        = {{xCPS}: an {xAct} package for covariant 
                  phase space, {N}oether charges, and entropy},
  year         = {2026},
  howpublished = {\href{https://github.com/juanmargalef/xCPS}
                  {\texttt{github.com/juanmargalef/xCPS}}},
  note         = {Mathematica package}
}

@misc{TInvar,
  author       = {Kiely, Kevin and Wardell, Barry and 
                  Ottewill, Adrian and 
                  Mart\'{\i}n-Garc\'{\i}a, Jos\'e M.},
  title        = {{TInvar}: Canonicalization of {R}iemann 
                  expressions with free indices},
  howpublished = {\href{https://xact.es/}{\texttt{xact.es}}},
  note         = {Mathematica package, part of the xAct bundle}
}
\end{document}